\documentclass[12pt,preprint]{aastex}
\usepackage{graphicx}
\usepackage{amsfonts,amsmath,amssymb,mathrsfs}
\usepackage{color}

\newcommand{\be}{\begin{eqnarray}}
\newcommand{\ee}{\end{eqnarray}}
\newcommand{\rar}{\rightarrow}

\begin{document}

\title{Testing the space-time geometry around black hole candidates with the available radio and X-ray data}

\author{Cosimo Bambi}
\affil{Center for Field Theory and Particle Physics \& Department of Physics,\\ 
Fudan University, 200433 Shanghai, China}
\email{bambi@fudan.edu.cn}

\date{\today}

\begin{abstract}
Astrophysical black hole candidates are thought to be the Kerr black holes
predicted by General Relativity, but the actual nature of these objects has
still to be proven. The Kerr black hole hypothesis can be tested by observing 
strong gravity features and check if they are in agreement with the predictions
of General Relativity. In particular, the study of the properties of the 
electromagnetic radiation emitted by the gas of the accretion disk can
provide information on the geometry of the space-time around these
objects and constrain possible deviations from the Kerr background.
\end{abstract}



\section{Introduction}

The predictions of General Relativity have been confirmed by experiments 
in the Earth's gravitational field \citep{llr,gpb}, by spacecraft missions in 
the Solar System \citep{cassini}, and by accurate radio observations of 
binary pulsars \citep{wt05,pulsar} (for a general review, see e.g. \citet{will}). 
In all these environments, the gravitational field is weak, in the sense that 
one can write $g_{tt} = 1 + \phi$ with $|\phi| \ll 1$. The validity of the theory 
in the regime of strong gravity, when the approximation $|\phi| \ll 1$ breaks 
down, is instead still unexplored. The ideal laboratory to test strong gravitational 
fields is the space-time around astrophysical black hole (BH) candidates.

In 4-dimensional General Relativity, uncharged BHs are described by the 
Kerr solution and are completely specified by two parameters: the mass, $M$, 
and the spin angular momentum, $J$ \citep{hair1,hair2}\footnote{The Kerr 
metric is a vacuum exact solution even in other theories of gravity, but its 
uniqueness is not guaranteed in general \citep{dimk}. The scenario of Kerr 
BHs in an alternative theory of gravity may be tested by looking for observational 
features that do not depend only on the metric of the background, such as
the quasi-normal modes of the compact object \citep{bs}.}. A 
fundamental limit for a Kerr BH is the bound $|a_*| \le 1$, where $a_* = J/M^2$ is 
the spin parameter. This is the condition for the existence of the event horizon: 
for $|a_*|>1$, the event horizon disappears, and the central singularity 
becomes naked, violating the weak cosmic censorship conjecture  \citep{wccc}.
Despite the apparent possibility of forming naked singularities from regular 
initial data (see e.g. \citet{jm1,jm2} and references therein), the existence of a 
Kerr naked singularity or of a super-spinning Kerr compact object (in which the
singularity may be replaced by an extremely compact object made of 
exotic matter) can be excluded for at least two reasons: it is apparently 
impossible to make a star collapse with $|a_*|>1$  \citep{lrezz} or overspin an 
already existing Kerr BH up to $|a_*|>1$ \citep{eb1,eb2}, and, even if created, 
this super-spinning object would be unstable and should quickly decay 
\citep{dotti,pani}.

Astrophysical BHs presumably form from the gravitational collapse of matter, 
while the BH uniqueness theorems require a stationary space-time. 
However, initial deviations from the Kerr solution should be quickly radiated 
away through the emission of gravitational waves \citep{price1,price2} (but 
see \citet{hair3}), and the geometry of the space-time around the compact
object should be well described by the Kerr metric\footnote{Deviations from
the Kerr metric due to the presence of an accretion disk should also be
extremely small in the case of a compact object in an X-ray binary system.
However, the presence of surrounding material might be important for some
super-massive compact objects in galactic nuclei.}. At the moment, there are at 
least two classes of astrophysical BH candidates (for a review, see e.g. \citet{n05}): 
stellar-mass compact objects in X-ray binary systems ($M \approx 5 - 20$~$M_\odot$), 
and super-massive bodies at the center of every normal galaxy ($M \sim 10^5 
- 10^9$~$M_\odot$). The measurement of the mass of the two classes of 
BH candidates is robust, because obtained by dynamical methods and without any 
assumption about the nature of these objects: basically, we can study the orbital 
motion of gas or of individual stars orbiting around the compact object and we can 
infer the mass of the latter by using Newtonian mechanics. Stellar-mass BH 
candidates are too heavy to be neutron or quark stars \citep{rr74,kb96}, while 
at least some of the super-massive objects in galactic nuclei are too heavy, 
compact, and old to be clusters of non-luminous bodies \citep{maoz}. 
The non-observation of electromagnetic radiation emitted by the possible
surface of these objects may also be interpreted as an indications that 
they have an event horizon and are therefore BHs \citep{hor1,hor2,hor3,hor4}
(see however \citet{a-hor} and \citet{cb-hor}). As there is no alternative 
explanation in the framework of conventional physics, astrophysical BH candidates 
are thought to be the Kerr BHs predicted by General Relativity, even if we have 
no evidence that the geometry of the space-time around them is really described 
by the Kerr solution.

The first proposal for testing the Kerr BH hypothesis was put forward by Ryan, 
who suggested to observe the gravitational waves emitted by an extreme-mass 
ratio inspiral (EMRI), i.e. a system consisting of a stellar-mass compact object 
orbiting around a super-massive BH candidate~\citep{ryan}: as future space-based 
gravitational wave detectors will be able to observe $\sim 10^4 - 10^6$ 
gravitational wave cycles emitted by an EMRI while the stellar-mass body is 
inspiraling into the gravitational field of the super-massive object, even a small 
deviation from the Kerr geometry will build up an observable dephasing in the 
gravitational waveforms, thus allowing one to map the space-time of the 
super-massive BH candidate with very high accuracy~\citep{gb06,bc07}. Besides 
tests based on gravitational waves, the space-time geometry around BH candidates 
can be probed by very accurate observations of the orbital motion of stars. In the 
case of a stellar-mass BH candidate in an X-ray binary system, that is possible 
if the companion star is a radio pulsar~\citep{wex}. In the case of the super-massive 
BH candidate at the center of our Galaxy, that might be achieved by monitoring 
stars orbiting at milliparsec distances from the compact object~\citep{will08}.

Recently, there has been an increasing new interest in the subject. While previous
studies focused on future facilities (like space-based gravitational wave detectors)
or possible future discoveries (like a BH binary with a pulsar companion),
more recent efforts deal with the possibility of testing the Kerr-nature of 
astrophysical BH candidates by studying the properties of the electromagnetic
radiation emitted by the gas of the accretion disk, for which there are already
available data. In the last 5-8 years, there have been significant progresses
in the understanding of the accretion process onto BH candidates and astronomers
have developed several approaches to estimate the spin parameter of these
objects under the assumption of Kerr background. One can typically extend
these techniques to constrain/discover possible deviations from the Kerr
metric \citep{cb-review}.

The content of the paper is as follows. In Section~\ref{s-th}, I review the 
theoretical framework to test the Kerr-nature of astrophysical BH candidates: 
the Novikov-Thorne model, which can be used to describe thin accretion disks 
in generic stationary and axisymmetric space-times, and the choice of a
non-Kerr background metric to parametrize possible deviations from the 
Kerr solution. In Section~\ref{s-ob}, I show some approaches to probe the
geometry of the space-time around astrophysical BH 
candidates from the analysis of the electromagnetic 
radiation emitted by the gas of accretion: the estimate of the radiative
efficiency of AGNs, which can provide a rough idea of the size of 
deformations that may already be considered unlikely, the
continuum-fitting method and the broad K$\alpha$ iron line analysis,
which are the two most popular techniques currently used by astronomers 
to estimate the spin parameter of these objects and can be naturally extended
to constrain possible deviations from the Kerr background, the high-frequency
quasi-periodic oscillations (QPOs) and the estimate of the jet power,
which may potentially be powerful probes, but in both cases the associated 
physical mechanism is not yet well understood from the theoretical point of view. 
Throughout the paper, I use units in which $G_{\rm N} = c = 1$, unless stated 
otherwise.

\section{Accretion process onto a compact object \label{s-th}}

\subsection{Novikov-Thorne model}

Geometrically thin and optically thick accretion disks are described by the
Novikov-Thorne (NT) model~\citep{nt1,nt2}, which is the relativistic generalization 
of the Shakura-Sunyaev model~\citep{ss-m}. The disk is thin in the sense 
that the disk opening angle is $h = H/r \ll 1$, where $H$ is the 
thickness of the disk at the radius $r$. In the Kerr background, there are 
four parameters (BH mass $M$, BH spin parameter $a_*$, mass accretion 
rate $\dot{M}$, and viscosity parameter $\alpha$), but the model can be 
easily extended to any (quasi-)stationary, axisymmetric, and asymptotically 
flat space-time. Accretion is possible because viscous magnetic/turbulent 
stresses and radiation transport energy and angular momentum outwards. 
The model assumes that the disk is on the equatorial plane and that the 
disk's gas moves on nearly geodesic circular orbits. For long-term accretions, 
the disk is adjusted on the equatorial plane as a result of the Bardeen-Petterson 
effect~\citep{bp} (such a mechanism works even in non-Kerr backgrounds~\citep{cb-s1}).
In the case of BH candidates in X-ray binaries, the disk is likely
on the equatorial plane if the two bodies formed from the collapse of the
same cloud \citep{tilt}.
The assumption of nearly geodesic circular orbits requires that the radial 
pressure is negligible compared to the gravitational force of the BH, which
is indeed the case for thin disks. Heat advection is ignored (it scales as 
$\sim h^2$) and energy is radiated from the disk surface. Magnetic fields 
are also ignored.

The key-ingredient of the NT model is that the inner edge of the disk is
at the innermost stable circular orbit (ISCO), where viscous stresses are
assumed to vanish. When the gas's particles reach the ISCO, they quickly
plunge into the BH, without emitting additional radiation. At first approximation,
the {\it total efficiency} of the accretion process is
\be\label{eq-eta}
\eta_{\rm tot} &=& 1 - E_{\rm ISCO} \, ,
\ee
where $E_{\rm ISCO}$ is the specific energy of the gas at the ISCO radius 
and depends uniquely on the background geometry (see Appendix). In general, 
the total power of the accretion process is converted into radiation and kinetic 
energy of jet/wind outflows, so we can write $\eta_{\rm tot} = \eta_{\rm rad} + 
\eta_{\rm kin}$, but in the NT model $\eta_{\rm kin} = 0$. 
$\eta_{\rm rad}$ is the {\it radiative efficiency} and can be inferred 
from the bolometric luminosity $L_{\rm bol}$ if the mass accretion rate is 
known: $L_{\rm bol} = \eta_{\rm rad} \dot{M}$.

How good is the NT model to describe the accretion disk around astrophysical
BH candidates? For non-magnetized and weakly-magnetized disks, there 
is a common consensus that the NT model describes correctly thin disks, $h \ll 1$,
when the viscosity parameter is small, $\alpha \ll 1$ \citep{ap}. A common criterion 
to select sources with thin disks is that the bolometric to Eddington luminosity 
ratio, $L_{\rm bol}/L_{\rm Edd}$, does not exceed 0.3 \citep{grs1915}. In the case 
of magnetized disks, the issue is open and controversial. The GRMHD simulations 
in \citet{p10} (see also \citet{p12}) show that the stress at the inner edge of the 
disk scales as $h$ and the authors conclude that the NT model with vanishing 
stress boundary condition is recovered for $h \rar 0$. The GRMHD simulations 
in \citet{nk} and \citet{nkh} show instead large stress at the inner edge of the 
disk even when $h \rar 0$; according to these authors, the NT model cannot 
describe magnetized disks, even when the disk is very thin. It is not clear if the 
disagreement between the two groups can be attributed to different configurations 
of the magnetic fields, different resolution of the simulations, or something else.

\subsection{Parametrizing deviations from the Kerr geometry}

In the weak-field limit, General Relativity has been successfully tested in many
situations. Solar System's experiments are conveniently discussed within the
parametrized post-Newtonian (PPN) formalism. One can see that the weak-field 
limit of a general static and spherically symmetric metric can be put in the form 
\be
ds^2 &=& - \left(1 - \frac{2M}{r} + 2\beta\frac{M^2}{r^2} + . . . \right) \, dt^2
+ \left(1 + 2\gamma\frac{M}{r} + . . . \right)(dx^2 + dy^2 + dz^2) \, ,
\ee
where $\beta$ and $\gamma$ are the PPN parameters. In General Relativity,
$\beta = \gamma = 1$, but in other metric theories of gravity that may not be 
true. Observations can measure $\beta$ and $\gamma$ and thus constrain
possible deviations from the predictions of General Relativity.

The natural extension of the PPN formalism to test the space-time geometry
around astrophysical BH candidates, where gravity is strong, 
is to write a general stationary, axisymmetric, 
and asymptotically flat metric in terms of the mass moments, ${\mathcal M}_l$, 
and of the current moments, ${\mathcal S}_l$, of the compact object \citep{ryan}. 
In the case of a Kerr BH, all the moments are locked to the mass and angular 
momentum by the following relation \citep{hansen}:
\be
{\mathcal M}_l + i {\mathcal S}_l = M \left(\frac{iJ}{M}\right)^l \, .
\ee
By measuring at least three non-trivial multiple moments of the space-time 
around an astrophysical BH candidate, one can test the Kerr-nature of the body. 
The advantage of a post-Newtonian approach is that we do not need specific 
assumptions about the nature of the BH candidate. The disadvantage is that 
we can only probe the metric relatively far from the compact object, where the 
contribution of higher-order moments can be neglected, and therefore we 
need very accurate measurements, which may not be possible in a realistic 
astrophysical situation.

In order to test the space-time geometry close to a BH candidate, we can 
consider a metric describing the gravitational field (in General Relativity 
or in an alternative theory of gravity) of a compact object characterized by
its mass, its spin angular momentum, and a set of deformation parameters 
measuring possible deviations from the Kerr solution. In this way, one can
compare observations and theoretical predictions in this more general
background and check if the observations require that all the deformation parameters must 
vanish, as necessary if the space-time geometry around the BH candidate is really
described by the Kerr metric. In the simplest case, we can consider just 
one deformation parameter, which should be able to describe the gravitational
field around a compact object either more oblate or more prolate (according
to the sign of the deformation parameter) 
than a Kerr BH. If our metric does it, it is not very important if
it is an exact solution of the vacuum Einstein's equations (in this case, we 
are actually considering the possibility the BH candidate is a compact object 
made of exotic matter) or not, in the sense that we eventually get the same
qualitative constraints \citep{cb-eta,cb-agn,zilong}. 
In this review paper, I will focus the discussion on the
metric proposed by Johannsen and Psaltis (JP) to describe the gravitational
field around non-Kerr BHs in putative alternative theories of gravity \citep{jp-m}.
In Boyer-Lindquist coordinates, the JP metric is given by the line element
\be\label{eq-jp}
ds^2 &=& - \left(1 - \frac{2 M r}{\Sigma}\right) (1 + h) \, dt^2 + \nonumber\\
&& + \frac{\Sigma (1 + h)}{\Delta + a^2 h \sin^2\theta } \, dr^2 + \Sigma 
\, d\theta^2 - \frac{4 a M r \sin^2\theta}{\Sigma} (1 + h) \, dt \, d\phi + \nonumber\\
&& + \left[\sin^2\theta \left(r^2 + a^2 + \frac{2 a^2 M r \sin^2\theta}{\Sigma} \right)
+ \frac{a^2 (\Sigma + 2 M r) \sin^4\theta}{\Sigma} h \right] d\phi^2 \, ,
\ee
where $a = a_* M $, $\Sigma = r^2 + a^2 \cos^2\theta$,
$\Delta = r^2 - 2 M r + a^2$, and
\be
h = \sum_{k = 0}^{\infty} \left(\epsilon_{2k}
+ \frac{M r}{\Sigma} \epsilon_{2k+1} \right)
\left(\frac{M^2}{\Sigma}\right)^k \, .
\ee
This metric has an infinite number of deformation parameters $\epsilon_i$ and
the Kerr solution is recovered when all the deformation parameters are set to
zero. However, in order to reproduce the correct Newtonian limit, we have to
impose $\epsilon_0 = \epsilon_1 = 0$, while $\epsilon_2$ is strongly
constrained by Solar System experiments~\citep{jp-m}. In what follows, I will 
examine only the simplest case in which $\epsilon_3 \neq 0$, while all the 
other deformation parameters are set to zero.

\begin{figure}
\begin{center}  
\includegraphics[type=pdf,ext=.pdf,read=.pdf,width=8cm]{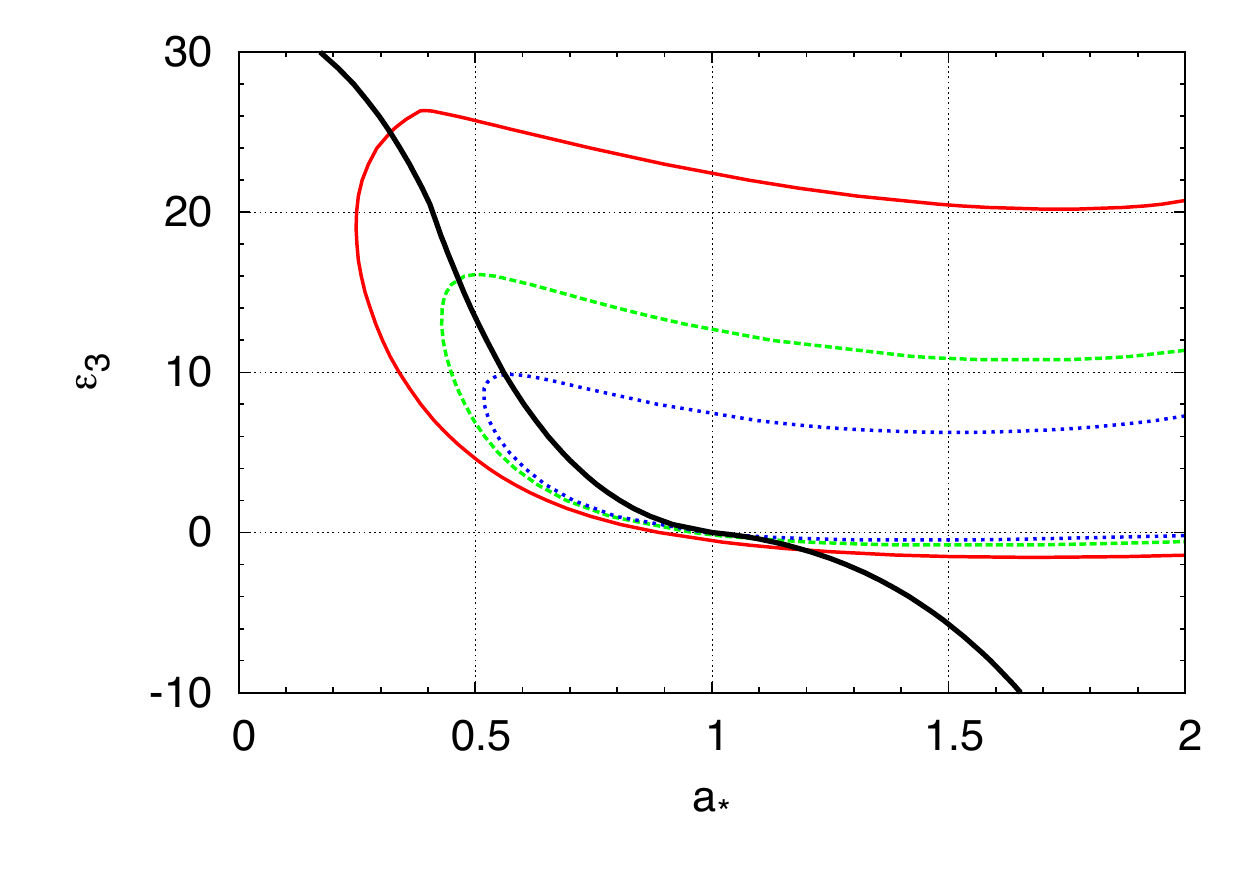}
\includegraphics[type=pdf,ext=.pdf,read=.pdf,width=8cm]{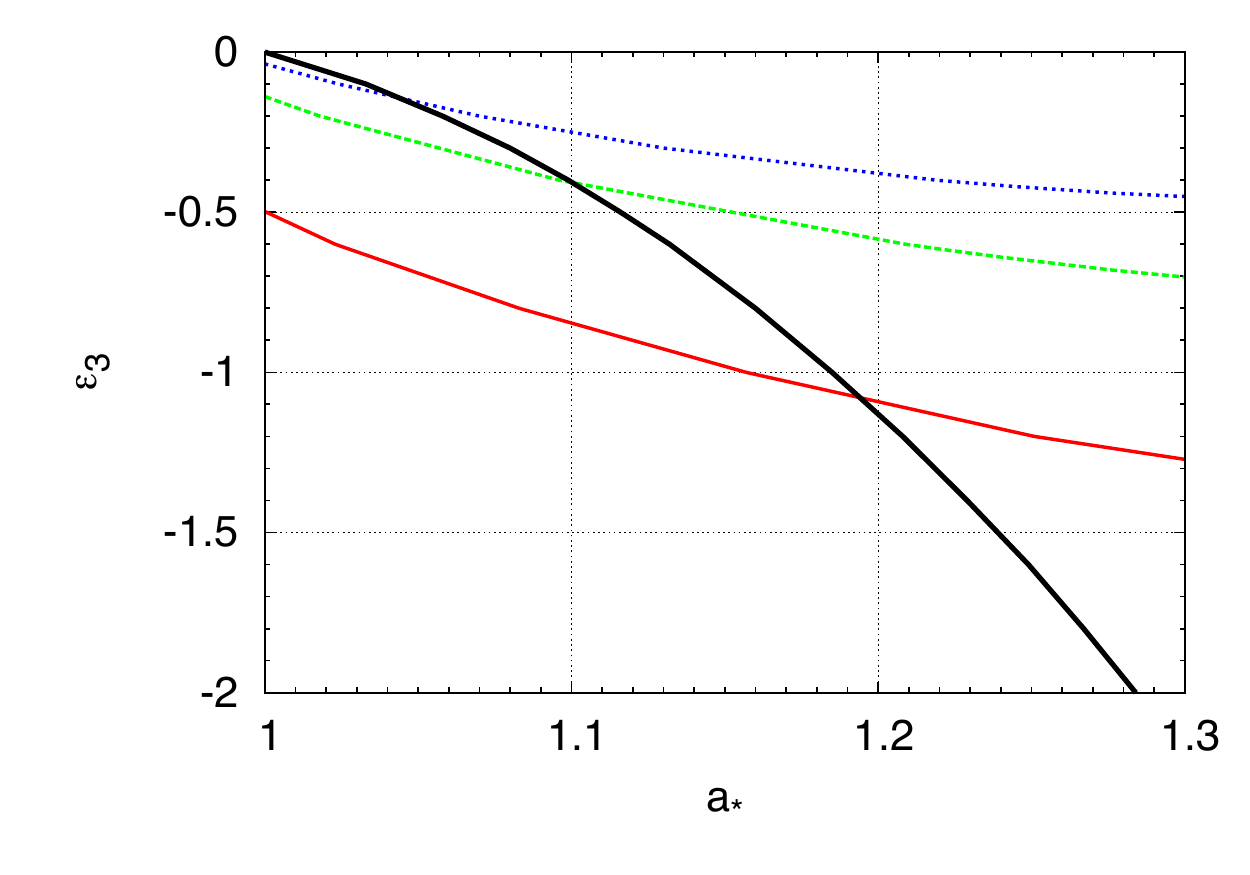}
\end{center}
\vspace{-0.6cm}
\caption{Total efficiency $\eta_{\rm tot} = 1 - E_{\rm ISCO}$ in the JP
background with spin parameter $a_*$ and deformation parameter 
$\epsilon_3$. $\eta_{\rm tot} = 0.15$ (red solid curve), 0.20 (green dashed curve), 
and 0.25 (blue dotted curve). 
The black solid curve is the equilibrium spin parameter $a_*^{\rm eq}$ 
obtained from Eq.~(\ref{eq-aeq}). The right panel is an enlargement of the 
parameter region $1.0 < a_* < 1.3$ and $-2.0 < \epsilon_3 < 0.0$.}
\label{f-eta}
\end{figure}

\section{Testing the Kerr-nature of BH candidates \label{s-ob}}

\subsection{Radiative efficiency of AGNs}

The value of the spin parameter of a compact object is determined
by the competition of three physical processes: the event creating
the object, mergers, and gas accretion. Accretion from a disk can 
potentially be a very efficient way to spin a compact object up. 
If the inner edge of the disk is at the ISCO radius, the gas's particles 
plunge into the compact object with specific energy $E_{\rm ISCO}$ 
and specific angular momentum $L_{\rm ISCO}$. The mass $M$ and 
the spin angular momentum $J$ of the compact object change respectively 
by $\delta M = E_{\rm ISCO} \delta m$ and $\delta J = L_{\rm ISCO} \delta m$, 
where $\delta m$ is the gas rest-mass. The evolution of the spin parameter 
turns out to be governed by the following equation~\citep{bard}
\be\label{eq-aeq}
\frac{da_*}{d\ln M} = \frac{1}{M} 
\frac{L_{\rm ISCO}}{E_{\rm ISCO}} - 2 a_* \, ,
\ee
neglecting the small effect of the radiation emitted by the disk and captured 
by the object. Prolonged disk accretion is a very efficient mechanism to 
spin the compact object up till an equilibrium spin parameter $a_*^{\rm eq}$, 
which is reached when the right-hand side of Eq.~(\ref{eq-aeq}) vanishes. For 
instance, an initially non-rotating Kerr BH reaches the equilibrium $a_*^{\rm eq} = 1$ 
after having increased its mass by a factor $\sqrt{6} \approx 2.4$~\citep{bard}.

One can argue that the most favorable scenario to produce fast-rotating 
super-massive objects at the center of galaxies is likely via prolonged disk 
accretion and that the maximum value for the spin parameter of these objects today
cannot exceed $a_*^{\rm eq}$ \citep{cb-eta}. The numerical value of $a_*^{\rm eq}$ 
depends on the metric of the space-time and it may exceed 1, as the bound
$|a_*| \le 1$ holds for Kerr BHs \citep{cb-s1,cb-s2,cb-l}. 
The black solid curve in Figure~\ref{f-eta}
shows the equilibrium spin parameter for the JP metric with
deformation parameter $\epsilon_3$. Objects on the left of the black solid curve 
have $a_* < a_*^{\rm eq}$ and the accretion process spins them up; objects on 
the right have $a_* > a_*^{\rm eq}$ and the accretion process spins them down.

In general, it is not easy to get an estimate of the radiative efficiency $\eta_{\rm rad}$, 
as the measurement of the mass accretion rate $\dot{M}$ is typically quite 
problematic. The mean radiative efficiency of AGNs can be inferred from the Soltan's 
argument \citep{soltan}, which relates the mass density of the super-massive BH 
candidates in the contemporary Universe with the energy density of the radiation
produced in the whole history of the Universe by the accretion process onto 
these objects. There are several sources of uncertainty in the final result, but a 
mean radiative efficiency $\eta_{\rm rad} > 0.15$ seems to be a conservative 
lower limit~\citep{elvis}. \citet{ho} find a mean radiative efficiency $\eta_{\rm rad} 
\approx 0.30 - 0.35$ without some important assumptions necessary in the 
original version of the Soltan's argument. \citet{davis} show how to estimate 
$\eta_{\rm rad}$ for individual AGNs and find that the most massive objects
have typically higher $\eta_{\rm rad}$, up to $\sim 0.3 - 0.4$.

It seems plausible to conclude that at least some of the super-massive BH 
candidates must have $\eta_{\rm rad} > 0.15$. While the accretion disk of
these objects may not be necessary thin, accretion from a thin disk is among
the most efficient ways to convert the rest-mass of the gas into radiation.
As $\eta_{\rm tot} \ge \eta_{\rm rad}$,
$E_{\rm ISCO} < 0.85$ and we can constrain deviations from the Kerr 
geometry \citep{cb-qagn,cb-eta}. A simplified explanation of how this is 
possible is the following. $E_{\rm ISCO}$ and $a_*^{\rm eq}$ increase/decrease 
if the gravitational force around the object increases/decreases, because the 
latter moves the ISCO to larger/smaller radii. Under the conservative assumption
that $\eta_{\rm tot} = \eta_{\rm rad}$, we can compute $\eta_{\rm tot}$ of the 
background metric and, combining this bound with the one for the maximum 
value of the spin parameter as a function of the deformation parameter, we get 
a well defined allowed region in the spin parameter-deformation parameter
plane. In the specific case of the JP metric with deformation parameter $\epsilon_3$,
we find the plot in Figure~\ref{f-eta}. While the specific value of the deformation
parameter depends on the choice of the metric of the background and a
physical interpretation is not straightforward, the maximum value for the spin 
parameter seems to depend only very weakly on the exact space-time, 
which makes this argument quite appealing \citep{cb-eta}. The maximum value
today of the spin parameter for the super-massive BH candidates in
galactic nuclei would be 1.19 if we require $\eta_{\rm tot} > 0.15$, 1.10
for $\eta_{\rm tot} > 0.20$, and 1.04 for $\eta_{\rm tot} > 0.25$.

More recently, \citet{zilong} have studied the accretion process from thick disks.
While in this case the details strongly depend on the unknown properties of the
accretion flow, at least in principle a thick disk can spin up a BH candidate more
efficiently. The evolution of the spin parameter is still given by Eq.~(\ref{eq-aeq}),
but the quantity $L_{\rm ISCO}/E_{\rm ISCO}$ is replaced by $L_{\rm in}/E_{\rm in}$,
where $E_{\rm in}$ and $L_{\rm in}$ are, respectively, the specific energy and the
specific angular momentum of the gas particles at the inner edge of the disk. For a 
thick disk, the inner edge of the disk can be inside the ISCO. So, if the super-massive BH 
candidates have experienced a period of super-Eddington accretion in the recent 
past, their spin parameter might exceed the equilibrium value $a_*^{\rm eq}$
computed from a thin disk. Under the most conservative assumptions, the
previous bounds on the maximum value of $a_*$ become slightly weaker: 1.30 
for $\eta_{\rm tot} > 0.15$, 1.18 for $\eta_{\rm tot} > 0.20$, and 1.10 for 
$\eta_{\rm tot} > 0.25$ \citep{zilong}.

\begin{figure}
\begin{center}  
\includegraphics[type=pdf,ext=.pdf,read=.pdf,width=8cm]{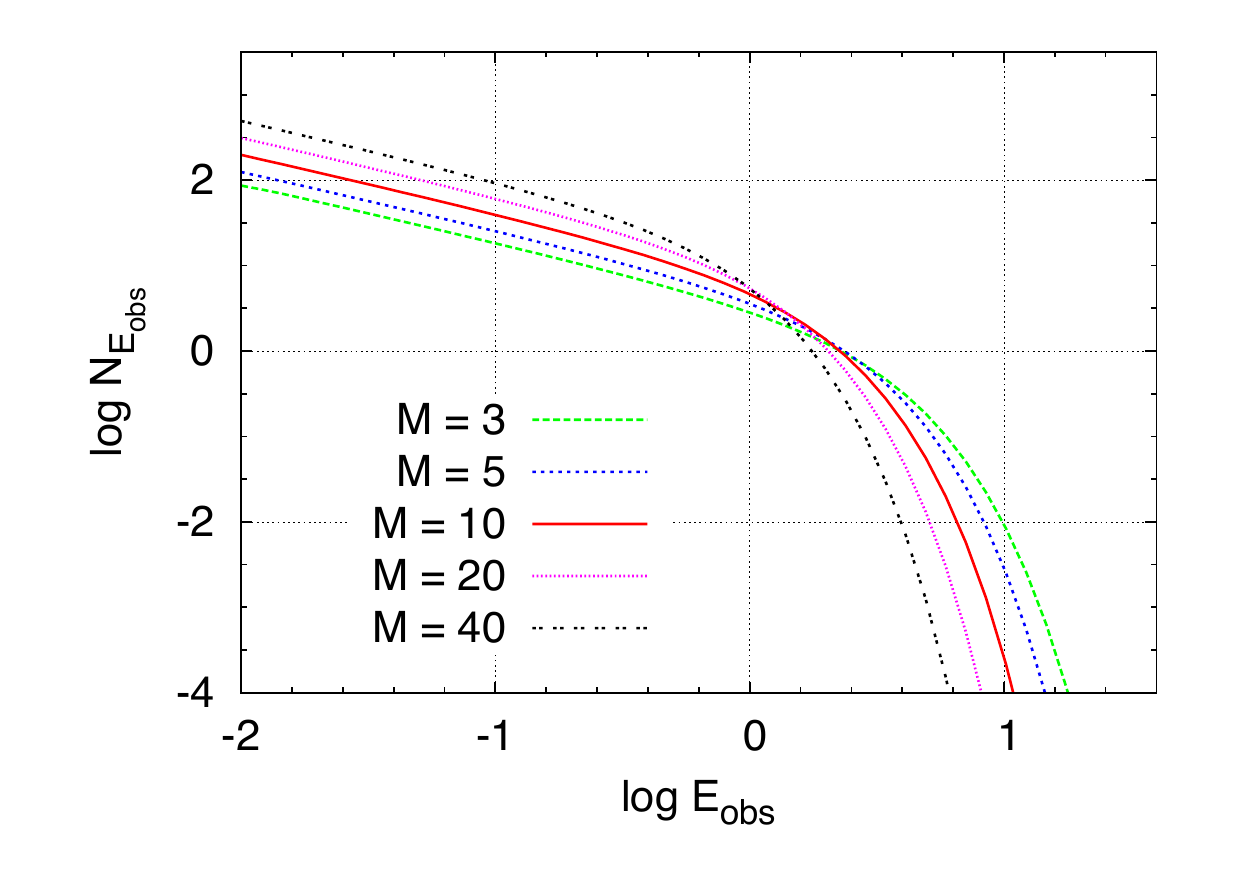}
\includegraphics[type=pdf,ext=.pdf,read=.pdf,width=8cm]{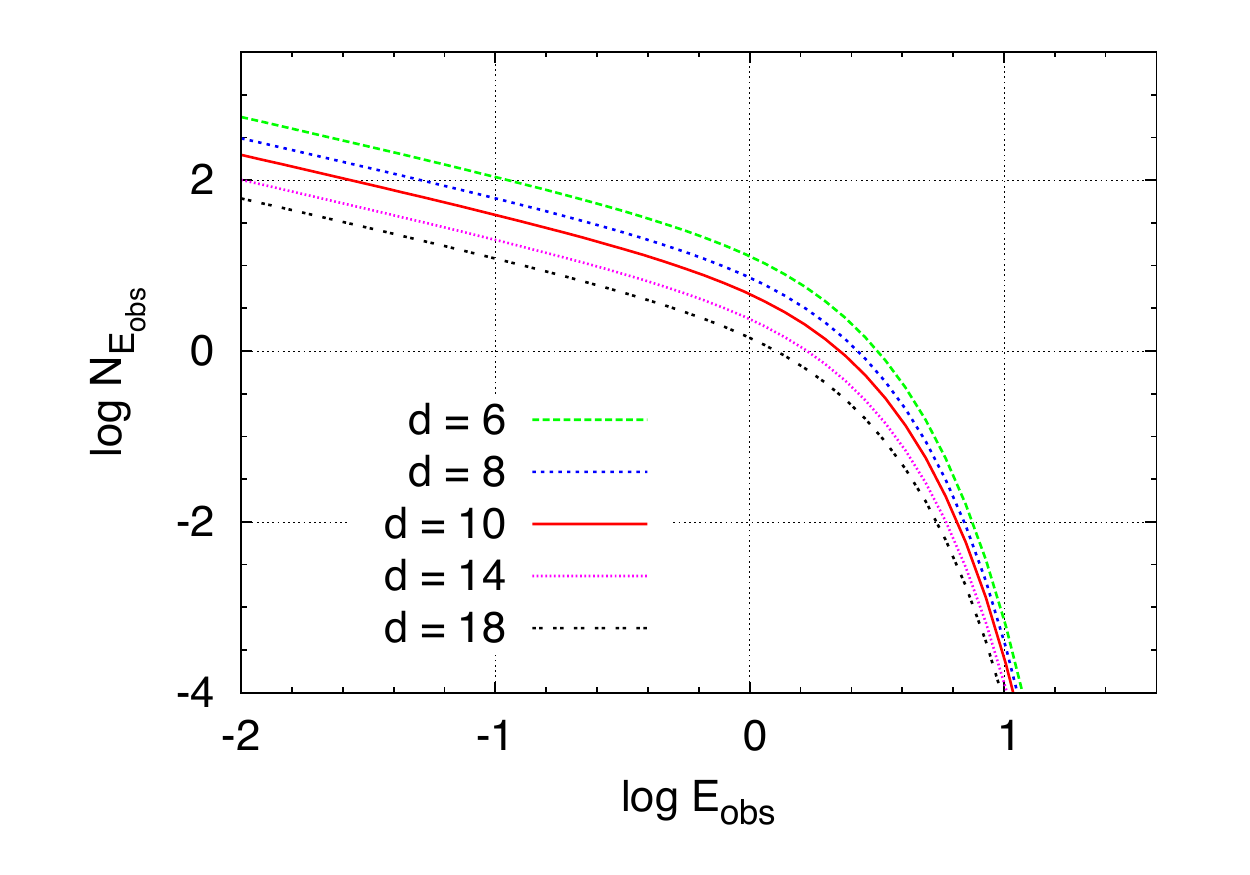} \\
\includegraphics[type=pdf,ext=.pdf,read=.pdf,width=8cm]{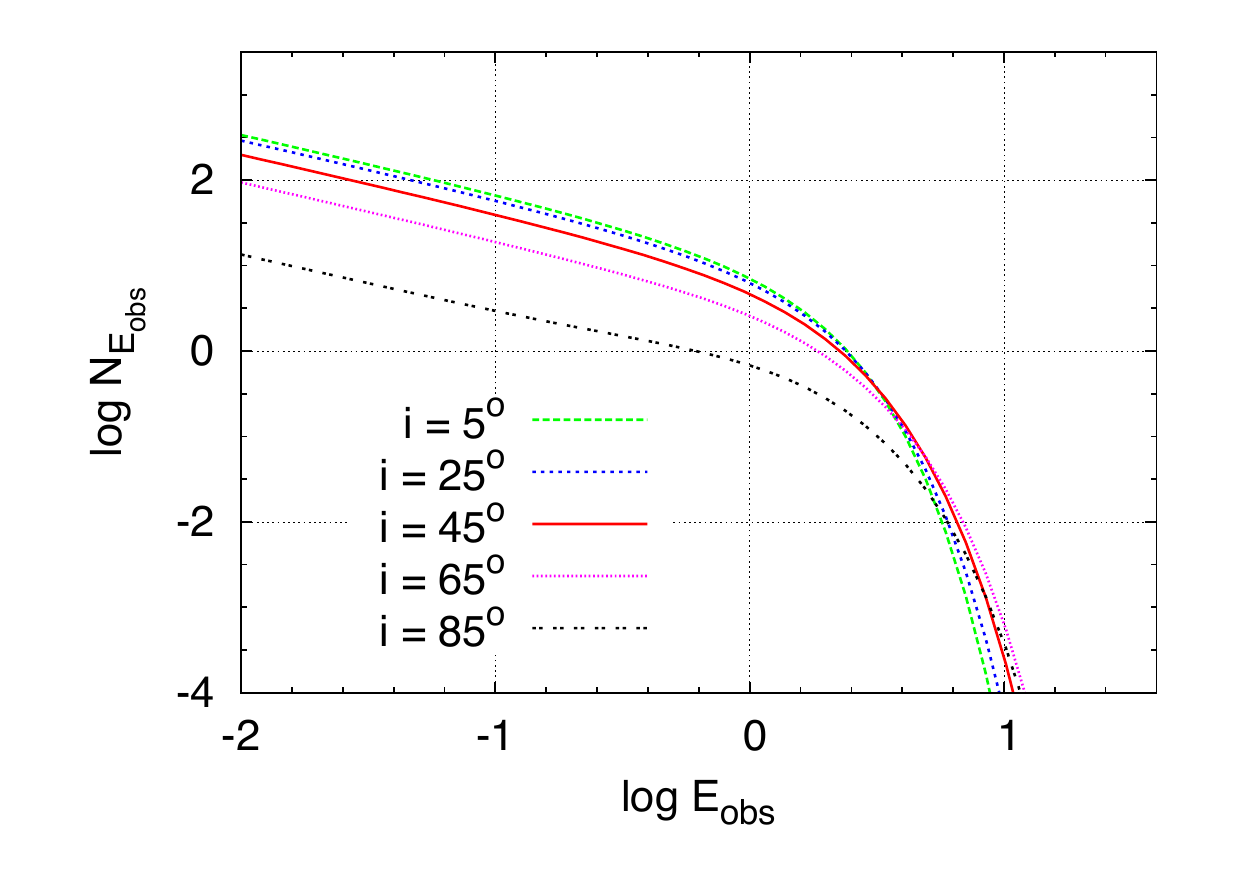}
\includegraphics[type=pdf,ext=.pdf,read=.pdf,width=8cm]{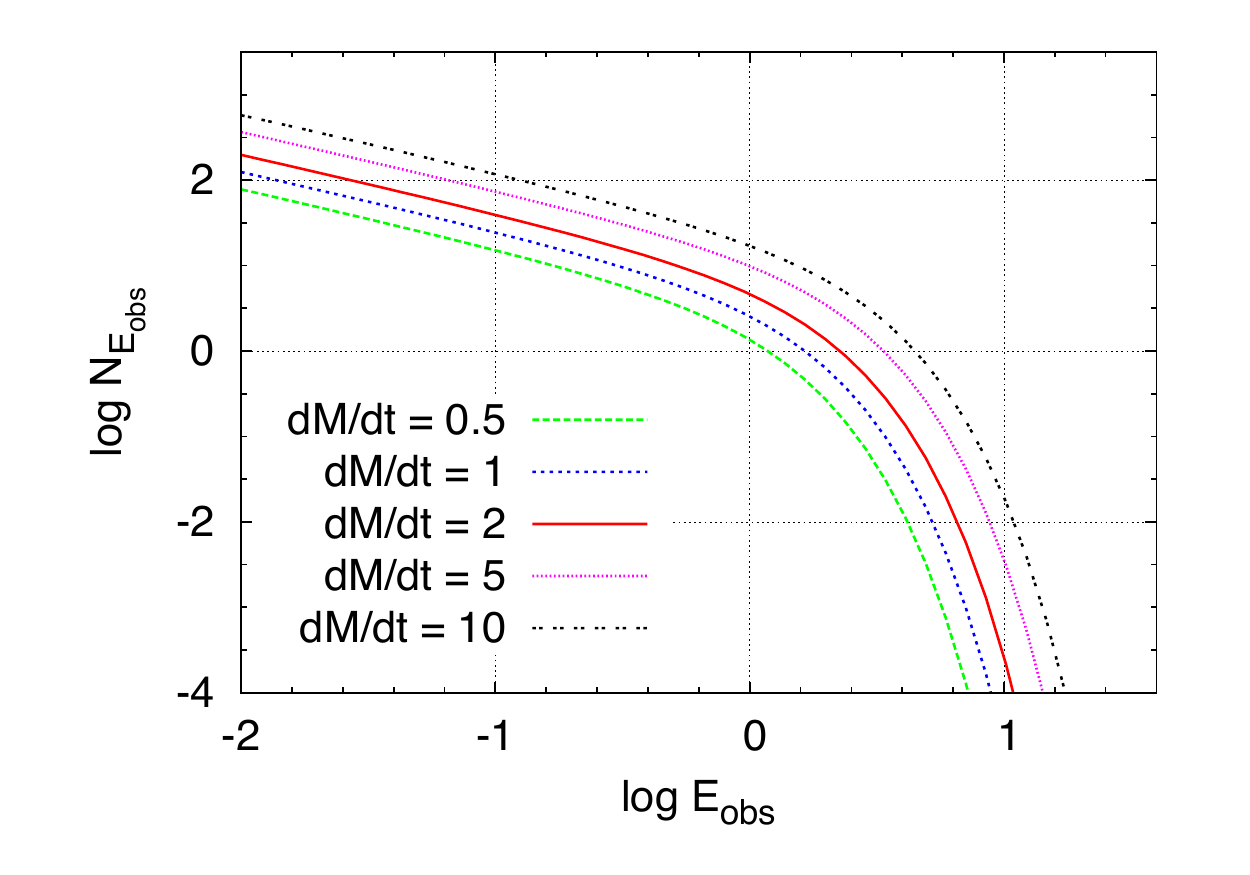} \\
\includegraphics[type=pdf,ext=.pdf,read=.pdf,width=8cm]{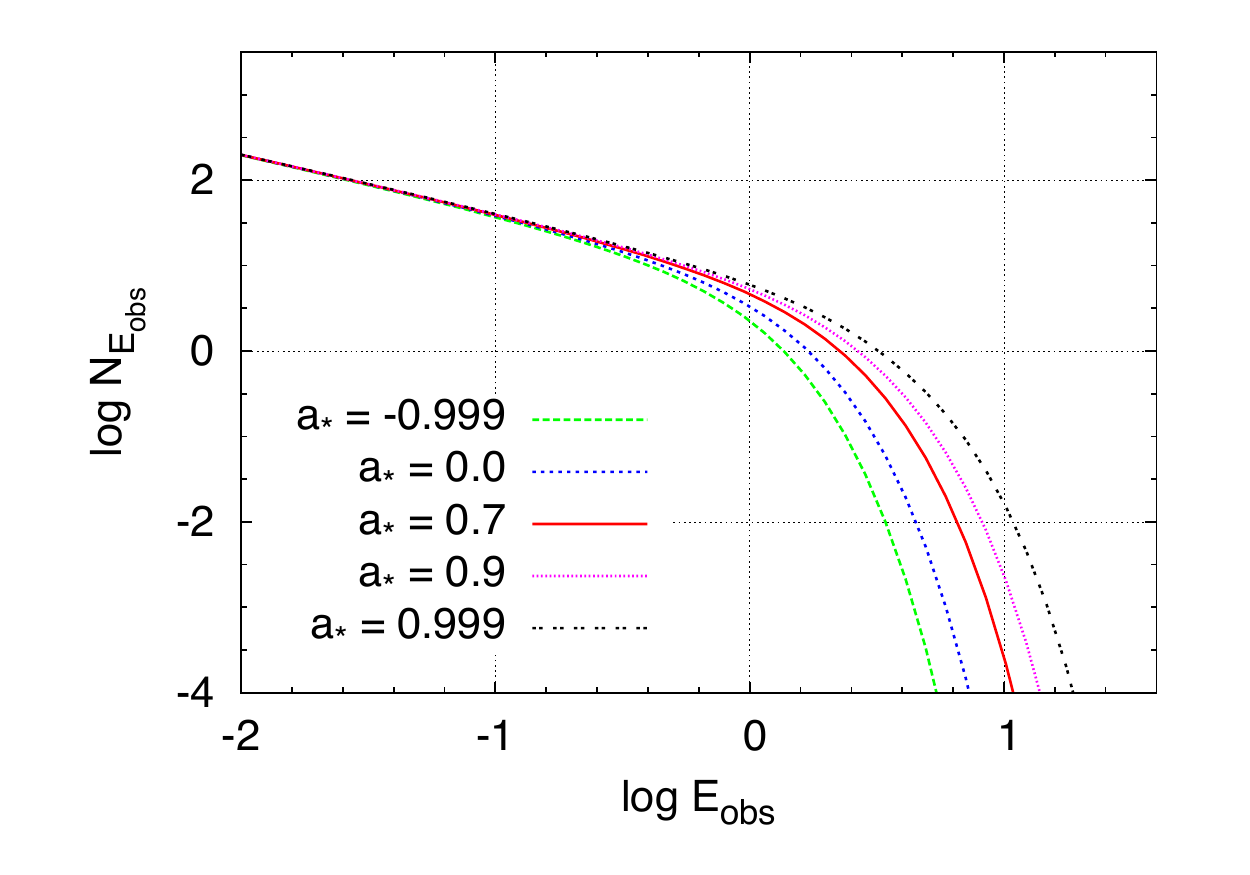}
\includegraphics[type=pdf,ext=.pdf,read=.pdf,width=8cm]{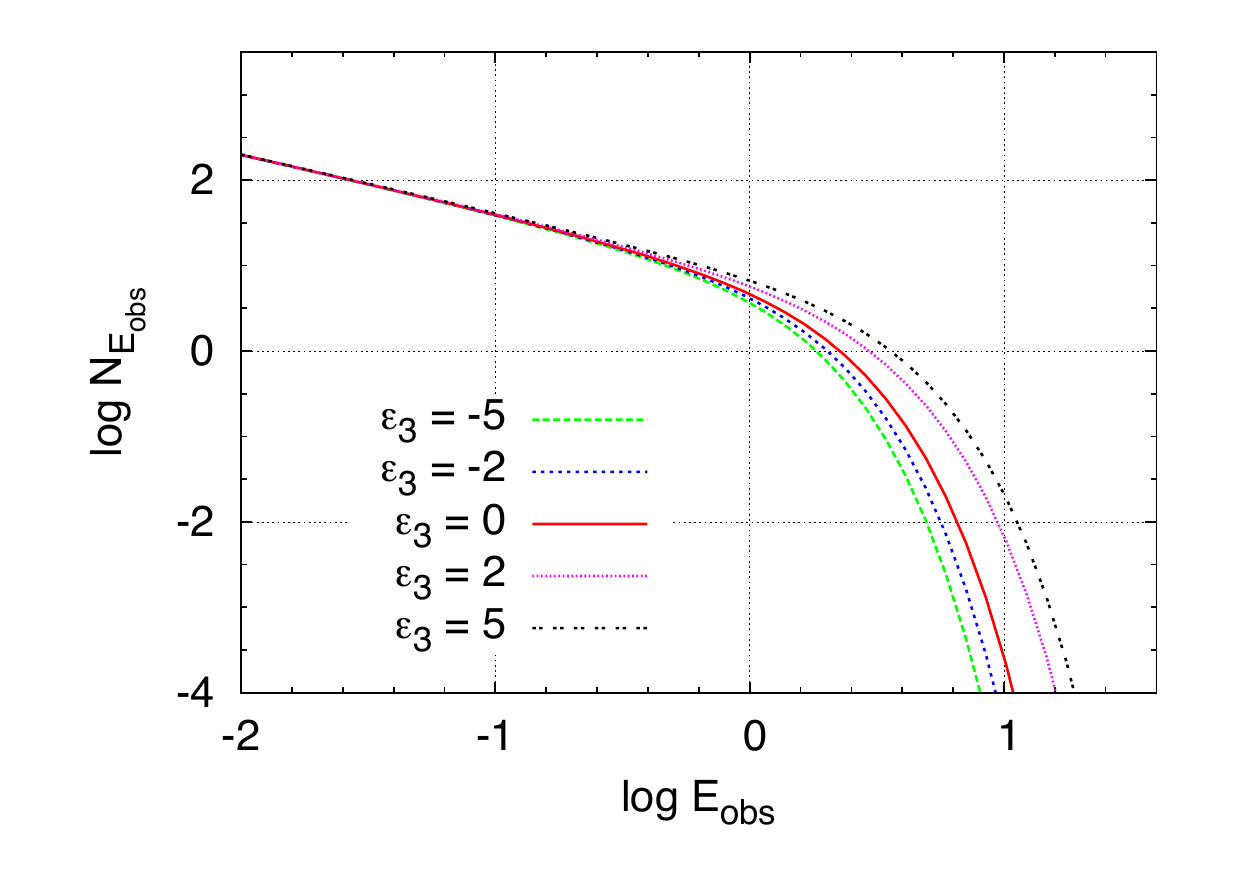}
\end{center}
\vspace{-0.6cm}
\caption{Thermal spectrum of a thin disk as a function of the parameters
of the model. Left top panel: mass of the BH candidate in units of $M_\odot$.
Right top panel: distance of the BH candidate in kpc. Left central panel:
inclination angle of the disk with respect to the line of sight of the distant
observer. Right central panel: mass accretion rate in units of $10^{18}$~g~s$^{-1}$.
Left bottom panel: spin parameter of the BH candidate. Right bottom
panel: deformation parameter of the BH candidate. When not shown,
the value of the parameters is: $M = 10$~$M_\odot$, $d = 10$~kpc, 
$i = 45^\circ$, $\dot{M} = 2 \cdot 10^{18}$~g~s$^{-1}$, $a_* = 0.7$, 
$\epsilon_3 = 0.0$. Flux density $N_{E_{\rm obs}}$ in 
$\gamma$~keV$^{-1}$~cm$^{-2}$~s$^{-1}$; photon energy $E_{\rm obs}$ 
in keV.}
\label{f-cfm_p1}
\end{figure}

\begin{figure}
\begin{center}  
\includegraphics[type=pdf,ext=.pdf,read=.pdf,width=8cm]{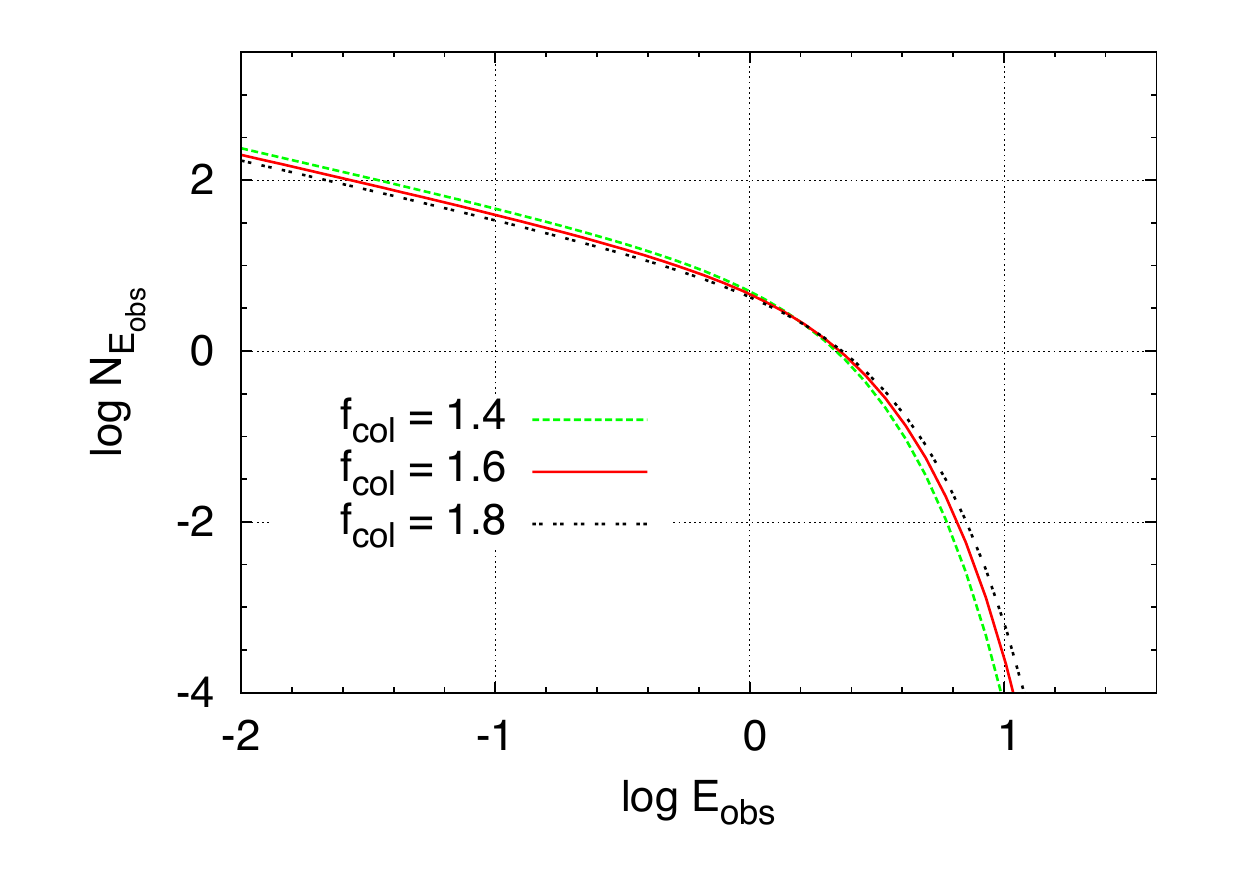}
\includegraphics[type=pdf,ext=.pdf,read=.pdf,width=8cm]{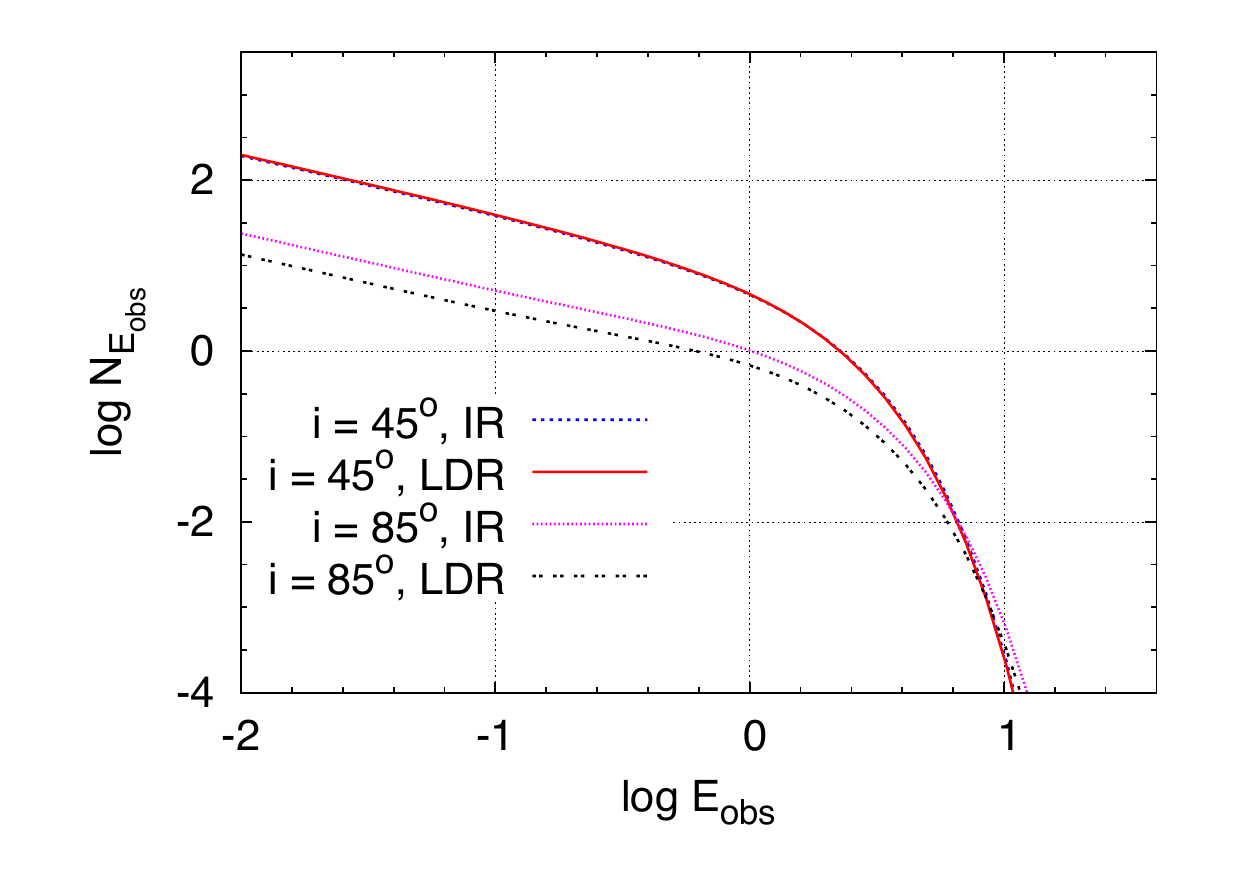}
\end{center}
\vspace{-0.6cm}
\caption{Left panel: effect of the color factor $f_{\rm col}$ on the thermal spectrum 
of a thin disk for $i = 45^\circ$. Right panel: effect of $\Upsilon$ on the thermal 
spectrum of a thin disk for $i = 45^\circ$ and $85^o$; IR/LDR = isotropic/limb-darkened
radiation (for $i = 45^\circ$, the two spectra are almost indistinguishable).
For both panels, the parameters of the model are: $M = 10$~$M_\odot$, 
$d = 10$~kpc, $\dot{M} = 2 \cdot 10^{18}$~g~s$^{-1}$, $a_* = 0.7$, 
$\epsilon_3 = 0.0$. Flux density $N_{E_{\rm obs}}$ in 
$\gamma$~keV$^{-1}$~cm$^{-2}$~s$^{-1}$; photon energy $E_{\rm obs}$ 
in keV.}
\label{f-cfm_p2}
\end{figure}

\begin{figure}
\begin{center}  
\includegraphics[type=pdf,ext=.pdf,read=.pdf,width=8cm]{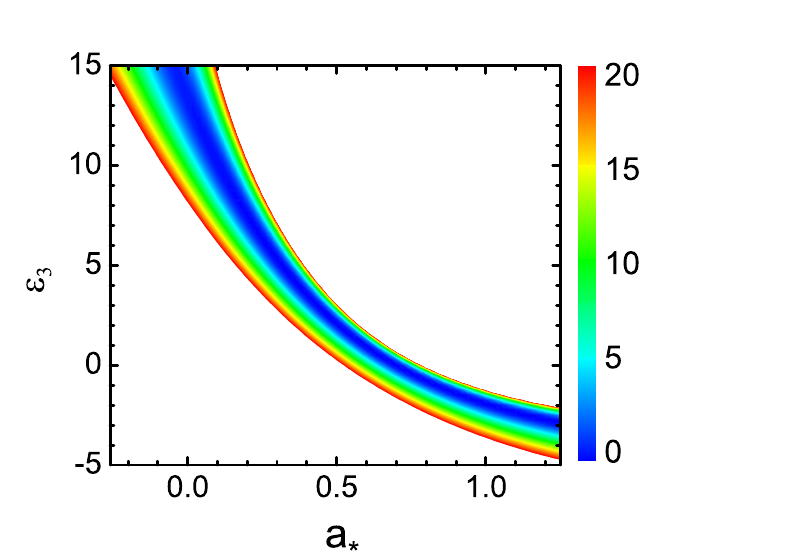}
\includegraphics[type=pdf,ext=.pdf,read=.pdf,width=8cm]{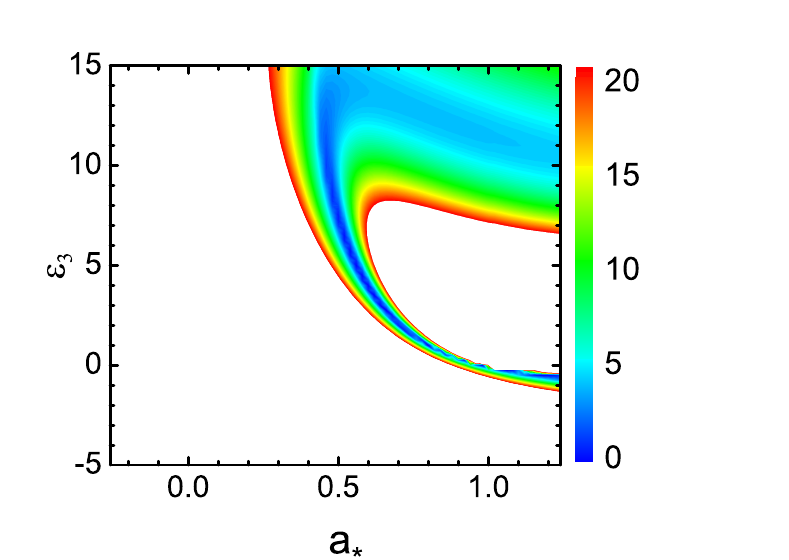}
\end{center}
\vspace{-0.6cm}
\caption{$\chi^2_{\rm red}$ from the comparison of the thermal spectrum of a 
thin accretion disk around a Kerr BH with spin parameter $\tilde{a}_*$ and a 
JP BH with spin parameter $a_*$ and deformation parameter $\epsilon_3$. 
Left panel: $\tilde{a}_* = 0.7$. Right panel: $\tilde{a}_* = 0.98$. See text for 
details.}
\label{f-cfm_fit}
\vspace{1.0cm}
\begin{center}  
\includegraphics[type=pdf,ext=.pdf,read=.pdf,width=8cm]{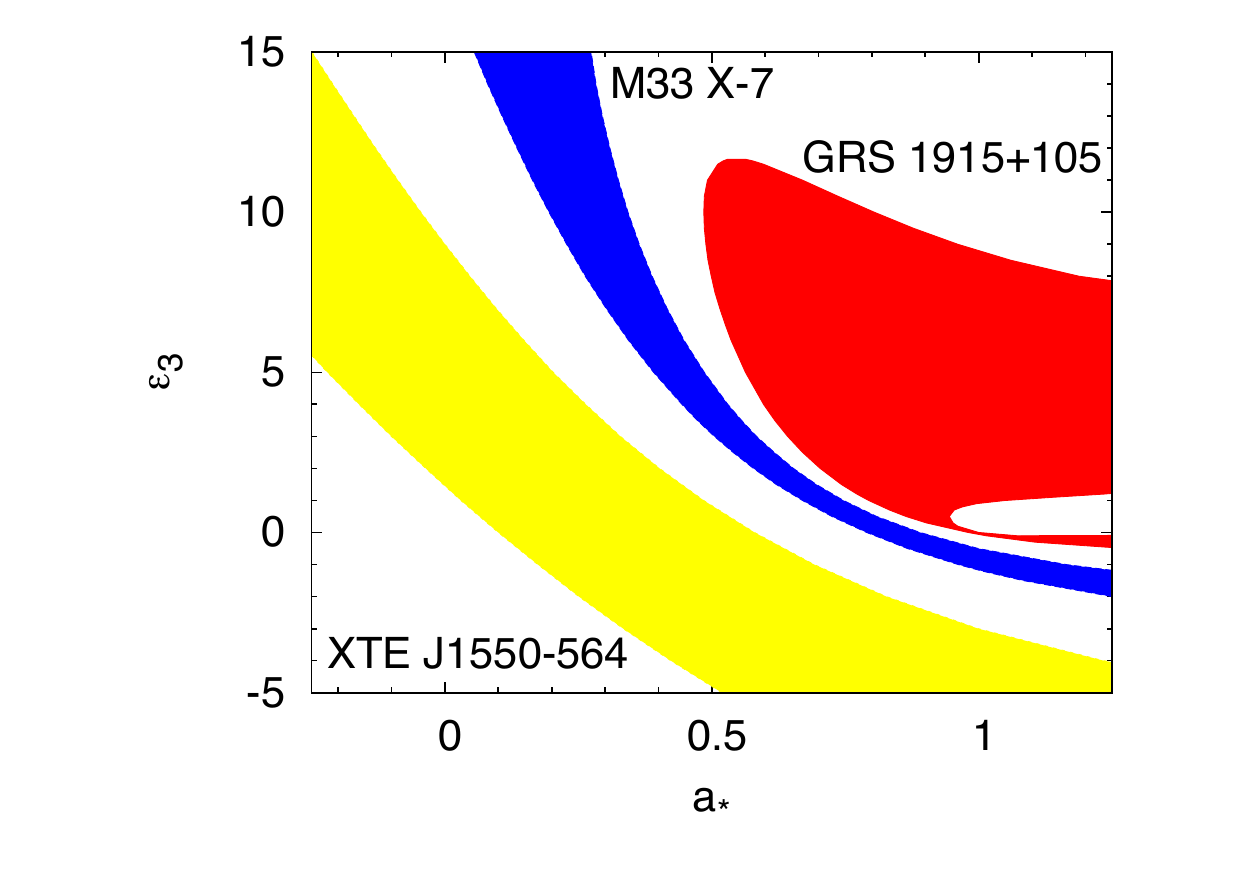}
\includegraphics[type=pdf,ext=.pdf,read=.pdf,width=8cm]{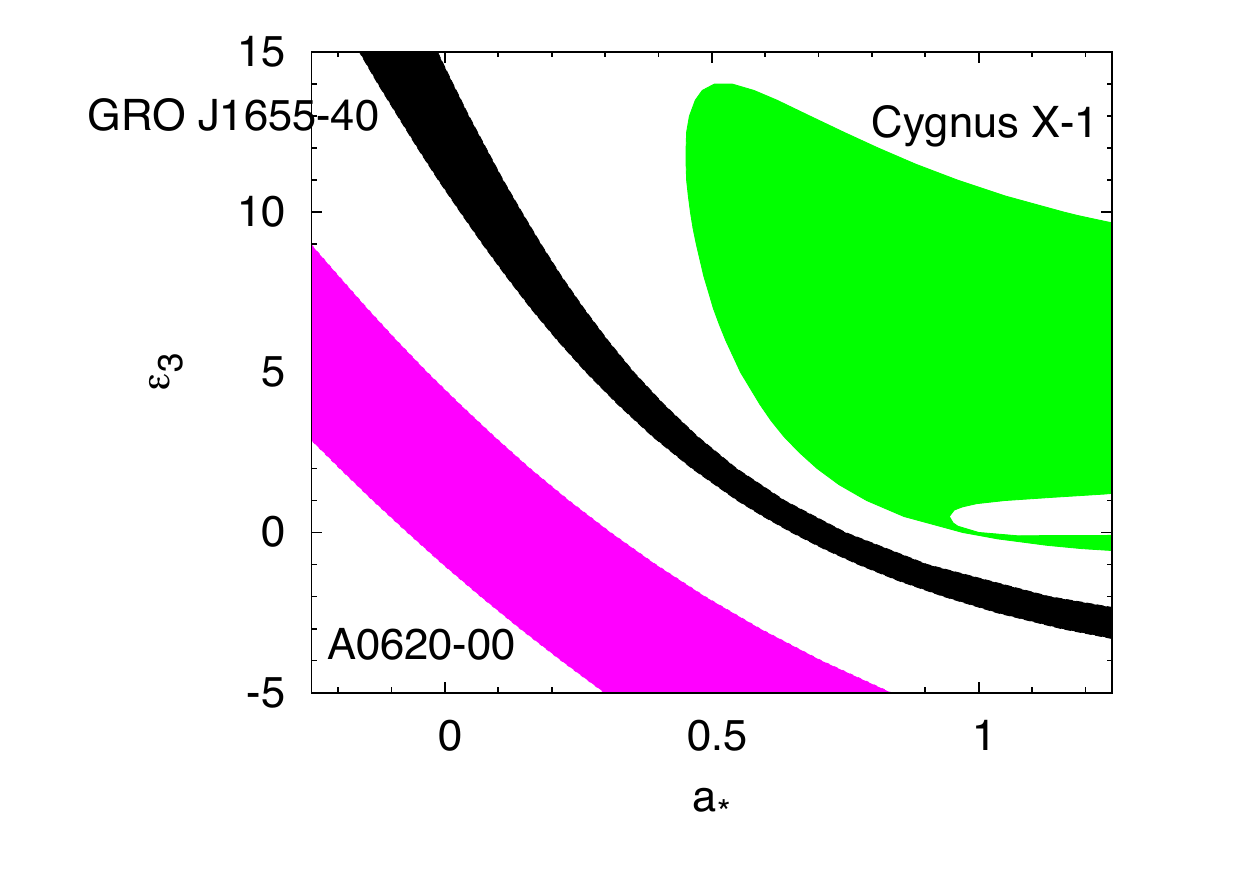}
\end{center}
\vspace{-0.6cm}
\caption{Constraints on the spin parameter-deformation parameter plane for
the BH candidates GRS~1915+105 (left panel, red area), M33~X-7 (left panel, 
blue area), XTE~J1550- (left panel, yellow area), Cygnus~X-1 (right panel,
green area), GRO~J1655-40 (right panel, black area), A0620-00 (right panel, 
magenta area). See text for details.}
\label{f-cfm_fit2}
\end{figure}

\subsection{Continuum-fitting method \label{s-cfm}}

In the NT model, the thermal spectrum of a geometrically thin 
and optically thick accretion disk around a Kerr BH observed far from the 
compact object depends on 5 parameters: 
BH mass $M$, BH distance $d$, inclination angle of the disk with respect to the
line of sight of the observer $i$, BH mass accretion rate $\dot{M}$, and BH spin 
$a_*$. If we can get independent measurements of $M$, $d$, and $i$, we can 
fit the data and estimate $\dot{M}$ and $a_*$. The technique is called 
continuum-fitting method \citep{z97,li05,cfm} and it can be applied only to 
stellar-mass BH candidates: the disk's temperature goes like $M^{-0.25}$,
and the peak of the spectrum is at about 1~keV for $M \sim 10$~$M_\odot$
and in the UV range for super-massive objects in galactic nuclei. In the latter 
case, a good measurement is impossible because of dust absorption. From 
the conservation laws for the rest-mass, angular momentum and energy, 
one can deduce three basic equations for the time-averaged radial structure 
of the disk. In particular, the time-averaged energy flux emitted from the 
surface of the disk is~\citep{nt2}
\be
\mathcal{F}(r) = \frac{\dot{M}}{4 \pi M^2} F(r) \, ,
\ee
where $F(r)$ is the dimensionless function
\be
F(r) = - \frac{\partial_r \Omega}{(E - \Omega L)^2} 
\frac{M^2}{\sqrt{-G}}
\int_{r_{\rm in}}^{r} (E - \Omega L) 
(\partial_\rho L) d\rho \, .
\nonumber\\
\ee
$E$, $L$, and $\Omega$ are, respectively, the conserved specific energy, 
the conserved axial-component of the specific angular momentum, and 
the angular velocity for equatorial circular geodesics (see Appendix); 
$G = - \alpha^2 g_{rr} g_{\phi\phi}$ 
is the determinant of the near equatorial plane metric, where $\alpha^2 =
g_{t\phi}^2/g_{\phi\phi} - g_{tt}$ is the lapse function; $r_{\rm in}$ is the 
inner radius of the accretion disk and in the NT model it is the ISCO radius.

Since the disk is in thermal equilibrium, the emission is blackbody-like 
and we can define an effective temperature $T_{\rm eff} (r)$ from the 
relation $\mathcal{F}(r) = \sigma T^4_{\rm eff}$, where $\sigma$ is the 
Stefan-Boltzmann constant. Actually, the disk's temperature near the inner 
edge of the disk can be high, up to $\sim 10^7$~K for stellar-mass BH candidates, 
and non-thermal effects are non-negligible. That is usually taken into account
by introducing the color factor (or hardening factor) $f_{\rm col}$. The
color temperature is $T_{\rm col} (r) = f_{\rm col} T_{\rm eff}$ and the
local specific intensity of the radiation emitted by the disk is
\be\label{eq-i-bb}
I_{\rm e}(E_{\rm e}) = \frac{2 E^3_{\rm e}}{h^2 c^2} \frac{1}{f_{\rm col}^4} 
\frac{\Upsilon}{\exp\left(\frac{E_{\rm e}}{k_{\rm B} T_{\rm col}}\right) - 1} \, ,
\ee
where $E_{\rm e}$ is the photon energy, $h$ is the Planck's constant, 
$c$ is the speed of light, $k_{\rm B}$ is the Boltzmann constant, and 
$\Upsilon$ is a function of the angle between the 3-momentum of the photon 
emitted by the disk and the normal of the disk surface, say $\xi$. The two 
most common options are $\Upsilon = 1$ (isotropic emission) and 
$\Upsilon = \frac{1}{2} + \frac{3}{4} \cos\xi$  (limb-darkened emission).

The calculation of the thermal spectrum of a thin accretion disk has
been extensively discussed in the literature; see e.g.~\citet{li05} 
and references therein for the Kerr case, and \citet{cb-cfm}, \citet{cn1} and 
\citet{cb-ray} for a background with generic deviations from the Kerr 
solution. The spectrum can be conveniently written in terms of 
the photon flux number density as measured by a distant observer, 
$N_{E_{\rm obs}}$. In the case of non-Kerr background, it is convenient 
to use a ray-tracing approach. The initial conditions $(t_0, r_0, \theta_0, \phi_0)$ 
for the photon with Cartesian coordinates $(X,Y)$ on the image plane of the 
distant observer are given by \citep{jp-rt}
\be
t_0 &=& 0 \, , \\
r_0 &=& \sqrt{X^2 + Y^2 + d^2} \, , \\
\theta_0 &=& \arccos \frac{Y \sin i + d \cos i}{\sqrt{X^2 + Y^2 + d^2}} \, , \\
\phi_0 &=& \arctan \frac{X}{d \sin i - Y \cos i} \, .
\ee
As the initial 3-momentum $\bf{k}_0$ must be perpendicular to the plane of the 
image of the observer, the initial conditions for the 4-momentum of the photon 
are
\be
k^r_0 &=& - \frac{d}{\sqrt{X^2 + Y^2 + d^2}} |\bf{k}_0| \, , \\
k^\theta_0 &=& \frac{\cos i - d \frac{Y \sin i + d 
\cos i}{X^2 + Y^2 + d^2}}{\sqrt{X^2 + (d \sin i - Y \cos i)^2}} |\bf{k}_0| \, , \\
k^\phi_0 &=& \frac{X \sin i}{X^2 + (d \sin i - Y \cos i)^2} |\bf{k}_0| \, , \\
k^t_0 &=& \sqrt{\left(k^r_0\right)^2 + r^2_0  \left(k^\theta_0\right)^2
+ r_0^2 \sin^2\theta_0  (k^\phi_0)^2} \, , 
\ee
where $k^t_0$ is inferred from the condition $g_{\mu\nu}k^\mu k^\nu = 0$ with
the metric tensor of a flat space-time (as the observer is located far from the 
compact object). The photon trajectory is numerically 
integrated backwards in time to the point of the photon emission on the 
accretion disk: in this way, we get the radial coordinate $r_{\rm e}$ at which 
the photon was emitted and the angle $\xi$ between the 3-momentum of the 
photon and the normal of the disk surface (necessary to compute $\Upsilon$).

The observerÕ's sky is divided into a number of small elements and the
ray-tracing procedure provides the observed flux density from each element; 
summing up all the elements, we get the total observed flux density of the 
disk. In the case of Kerr background, one can actually exploit the special 
properties of the Kerr solution and solve a simplified set of differential 
equations. That is not possible in a generic non-Kerr background and
one has to solve the second-order photon geodesic equations of the space-time. 
The photon flux number density is given by 
\be\label{eq-n2}
N_{E_{\rm obs}} &=&
\frac{1}{E_{\rm obs}} \int I_{\rm obs}(E_{\rm obs}) d \Omega_{\rm obs} = 
\frac{1}{E_{\rm obs}} \int g^3 I_{\rm e}(E_{\rm e}) d \Omega_{\rm obs} = 
\nonumber\\ &=& 
A_1 \left(\frac{E_{\rm obs}}{\rm keV}\right)^2
\int \frac{1}{M^2} \frac{\Upsilon dXdY}{\exp\left[\frac{A_2}{g F^{1/4}} 
\left(\frac{E_{\rm obs}}{\rm keV}\right)\right] - 1} \, ,
\ee
where $I_{\rm obs}$ and $E_{\rm obs}$ are, respectively, the specific
intensity of the radiation and the photon energy measured
by the distant observer. $d\Omega_{\rm obs} = dX dY / d^2$ 
is the element of the solid angle subtended by the image of the disk on the 
observer's sky. $g$ is the redshift factor
\be\label{eq-red}
g = \frac{E_{\rm obs}}{E_{\rm e}} = 
\frac{k_\alpha u^{\alpha}_{\rm obs}}{k_\beta u^{\beta}_{\rm e}}\, ,
\ee
where $k^\alpha$ is the 4-momentum of the photon, $u^{\alpha}_{\rm obs} = (-1,0,0,0)$ 
is the 4-velocity of the distant observer, and $u^{\alpha}_{\rm e} = (u^t_{\rm e},0,0,
\Omega u^t_{\rm e})$ is the 4-velocity of the emitter.  
$I_{\rm e}(E_{\rm e})/E_{\rm e}^3 = I_{\rm obs} (E_{\rm obs})/E_{\rm obs}^3$ follows 
from the Liouville's theorem. $A_1$ and $A_2$ are 
given by (for the sake of clarity, here I show explicitly $G_{\rm N}$ and $c$)
\be
A_1 &=&  
\frac{2 \left({\rm keV}\right)^2}{f_{\rm col}^4} 
\left(\frac{G_{\rm N} M}{c^3 h d}\right)^2 = 
\nonumber\\ &=& 
\frac{0.07205}{f_{\rm col}^4} 
\left(\frac{M}{M_\odot}\right)^2 
\left(\frac{\rm kpc}{d}\right)^2 \, 
{\rm \gamma \, keV^{-1} \, cm^{-2} \, s^{-1}} \, , \nonumber\\
A_2 &=&  
\left(\frac{\rm keV}{k_{\rm B} f_{\rm col}}\right) 
\left(\frac{G_{\rm N} M}{c^3}\right)^{1/2}
\left(\frac{4 \pi \sigma}{\dot{M}}\right)^{1/4} = 
\nonumber\\ &=& 
\frac{0.1331}{f_{\rm col}} 
\left(\frac{\rm 10^{18} \, g \, s^{-1}}{\dot{M}}\right)^{1/4}
\left(\frac{M}{M_\odot}\right)^{1/2} \, .
\ee
Using the normalization condition
$g_{\mu\nu}u^{\mu}_{\rm e}u^{\nu}_{\rm e} = -1$, one finds
\be
u^t_{\rm e} = - \frac{1}{\sqrt{-g_{tt} - 2 g_{t\phi} \Omega - g_{\phi\phi} \Omega^2}} \, ,
\ee
and therefore
\be\label{eq-red-g}
g = \frac{\sqrt{-g_{tt} - 2 g_{t\phi} \Omega - g_{\phi\phi} \Omega^2}}{1 + 
\lambda \Omega} \, ,
\ee
where $\lambda = k_\phi/k_t$ is a constant of the motion along the photon path
and can be evaluated from the photon initial conditions 
($\lambda = k_\phi/k_t = r_0 \sin\theta_0 k^\phi_0 / k^t_0$).

Figure~\ref{f-cfm_p1} shows the thermal spectrum of a thin disk as a function 
of the parameters of the model. $M$, $d$, and $i$ should be inferred from
independent measurements, while $a_*$, $\dot{M}$, and $\epsilon_3$
can be obtained by fitting the disk's spectrum. It is evident that variations
of $a_*$ and $\epsilon_3$ produce quite similar effects on the shape of
the spectrum, while $\dot{M}$ changes the spectrum in a very different way.
Figure~\ref{f-cfm_p2} shows the effects of variations of the color factor 
$f_{\rm col}$ (typically one should expect $f_{\rm col} = 1.6 - 1.7$)
and of the function $\Upsilon$ (whose effect is relevant only for large 
inclination angles).

In order to figure out how the analysis of the thermal spectrum of a thin disk
can constrain the geometry of the space-time around a BH candidate, we can 
compare the spectra of a Kerr BH and of a JP BH with deformation parameter 
$\epsilon_3$. We define the reduced $\chi^2$ as follows
\be\label{eq-chi2}
\chi^2_{\rm red} (a_*, \epsilon_3, \dot{M}) = 
\frac{\chi^2}{n} = \frac{1}{n} \sum_{i = 1}^{n} 
\frac{\left[N_i^{\rm JP} (a_*, \epsilon_3, \dot{M}) 
- N_i^{\rm Kerr} (\tilde{a}_*, \tilde{\dot{M}}) 
\right]^2}{\sigma^2_i} \, ,
\ee 
where the summation is performed over $n$ sampling energies $E_i$,
$N_i^{\rm JP}$ and $N_i^{\rm Kerr}$ are the photon flux number densities
respectively for the JP and Kerr metric, and $\sigma_i$ is the uncertainty.
As variations of $\dot{M}$ produce quite different effects on the shape of
the spectrum with respect to $a_*$ and $\epsilon_3$, its determination
is not correlated to the other two parameters and here we directly assume
$\dot{M} = \tilde{\dot{M}}$. Figure~\ref{f-cfm_fit} shows $\chi^2_{\rm red}$
for two cases: $\tilde{a}_* = 0.7$ (left panel) and $\tilde{a}_* = 0.98$
(right panel), assuming that the uncertainty is 10\% the 
photon flux number density; that is, $\sigma_i = 0.1 \, N_i^{\rm Kerr}$.
The continuum-fitting method basically measures the radiative efficency
$\eta_{\rm rad} = 1 - E_{\rm ISCO}$,
not the spin parameter $a_*$. In the Kerr metric, $\eta_{\rm rad}$ depends
only on $a_*$ and therefore one can infer the spin of the compact object.
In a generic non-Kerr background, one can only measure a combination
of $a_*$ and of the deformation parameters. The technique is indeed
unable to distinguish a Kerr BH with spin parameter $\tilde{a}_*$
from a very deformed object with completely different spin parameter but
very similar radiative efficiency \citep{cb-cfm}. An analysis of real data from 
specific sources will be presented in \citet{cb-mj}.

For the time being, the continuum-fitting method has been used to estimate
the spin parameter of 9 stellar-mass BH candidates under the assumption
that these objects are Kerr BHs. The list of these objects is reported in 
Table~\ref{tab1}. There are 2 objects that look like very fast-rotating Kerr
BHs (GRS~1915+105 and Cygnus~X-1), 4 objects that seem to be Kerr
BHs with a mid value of the spin parameter (LMC~X-1, M33~X-7, 4U~1543-47,
and GRO~J1655-40), and 3 objects whose data are consistent with
slow-rotating Kerr BHs (XTE~J1550-564, LMC~X-3, and A0620-00).
The measure of the spin parameter $a_*$ under the Kerr BH hypothesis
can be quickly translated into a measurement of $\eta_{\rm rad}$ and then
into an allowed region on the spin parameter-deformation
parameter plane, see Fig.~\ref{f-cfm_fit2}. The comparison of the
allowed region for GRO~J1655-40 (black area, right panel in Figure~\ref{f-cfm_fit2}) 
with the $\chi^2_{\rm red}$ in the left panel of Figure~\ref{f-cfm_fit}
shows that the continuum-fitting method measures indeed the 
radiative efficiency. For high values of the spin 
and of the deformation parameters, the rule of the radiative efficiency does
not work so well (see the case of GRS~1915+105 or Cygnus~X-1 and compare
with the allowed region in the right panel of Figure~\ref{f-cfm_fit}), but such a 
region of the plane is more likely unphysical, as it is probably impossible to
spin these very prolate compact objects to so high values of $a_*$ 
(see e.g. the black line in Figure~\ref{f-eta}).

\begin{table}
\centering
\begin{tabular}{ccccccccc}
\hline \hline
BH Binary & \hspace{0.5cm} & $a_*^{\rm Kerr}$ & $\eta_{\rm min}$ & $\eta_{\rm max}$ & \hspace{0.5cm} & Reference \\
\hline \hline
GRS~1915+105 && $a_* > 0.98$ & 0.234 & 0.423 && \citet{grs1915} \\
Cygnus~X-1 && $a_* > 0.97$ & 0.215 & 0.423 && \citet{cyg} \\
\hline
LMC~X-1 && $0.92 \pm 0.06$ & 0.139 & 0.234 && \citet{lmcx1} \\
M33~X-7 && $0.84 \pm 0.05$ & 0.120 & 0.151 && \citet{m33,m33e} \\
4U~1543-47 && $0.80 \pm 0.05$ & 0.112 & 0.136 && \citet{gro1655} \\
GRO~J1655-40 && $0.70 \pm 0.05$ & 0.097 & 0.112 && \citet{gro1655} \\
\hline
XTE~J1550-564 && $0.34 \pm 0.24$ & 0.0606 & 0.0892 && \citet{xte1550} \\
LMC~X-3 && $a_* < 0.3$ & 0.0365 & 0.0694 && \citet{lmcx3} \\
A0620-00 && $0.12 \pm 0.19$ & 0.0550 & 0.0699 && \citet{a0620} \\
\hline \hline
\end{tabular}
\caption{Continuum-fitting method results from the Harvard-Smithsonian CfA group. 
See references in the last column for more details.
\label{tab1}}
\end{table}

\begin{figure}
\begin{center}  
\includegraphics[type=pdf,ext=.pdf,read=.pdf,width=8cm]{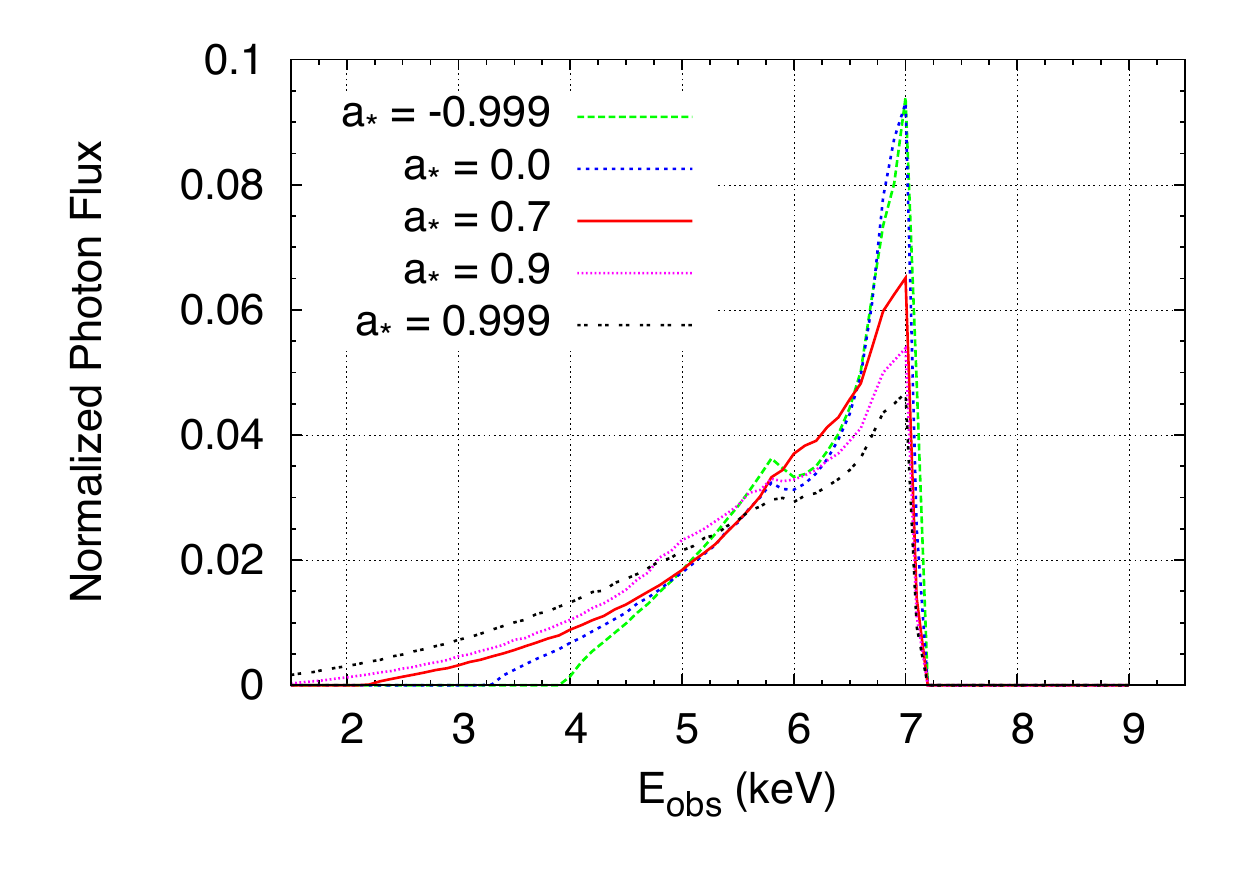}
\includegraphics[type=pdf,ext=.pdf,read=.pdf,width=8cm]{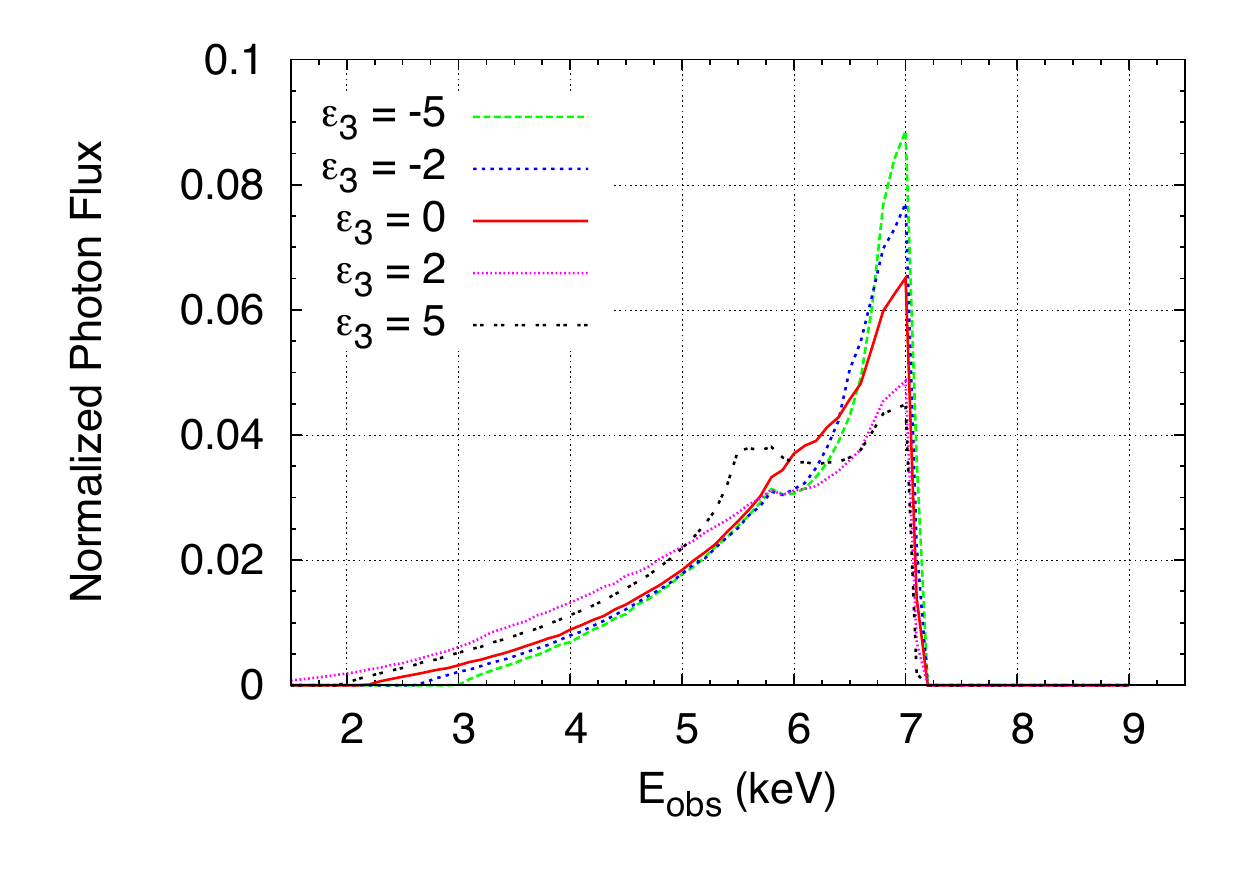} \\
\includegraphics[type=pdf,ext=.pdf,read=.pdf,width=8cm]{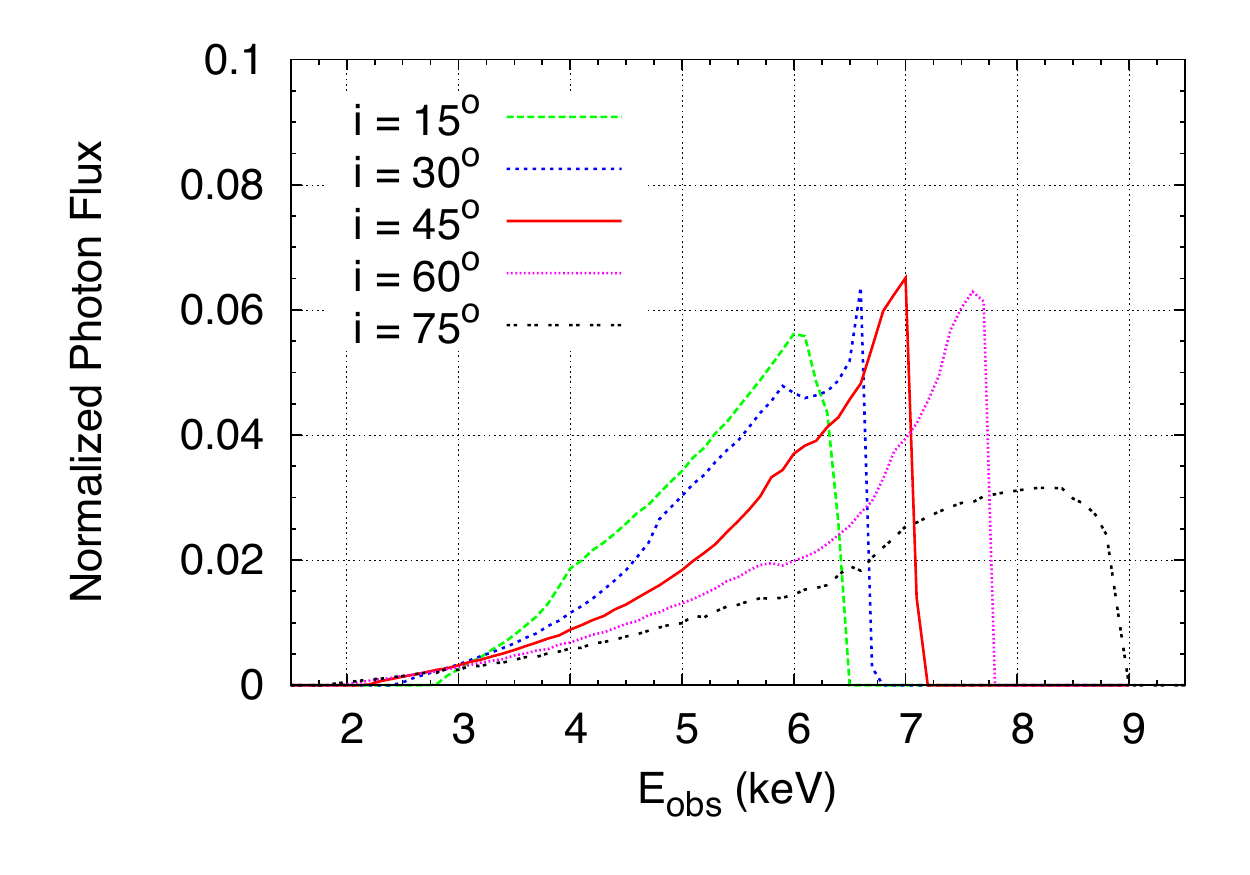} 
\includegraphics[type=pdf,ext=.pdf,read=.pdf,width=8cm]{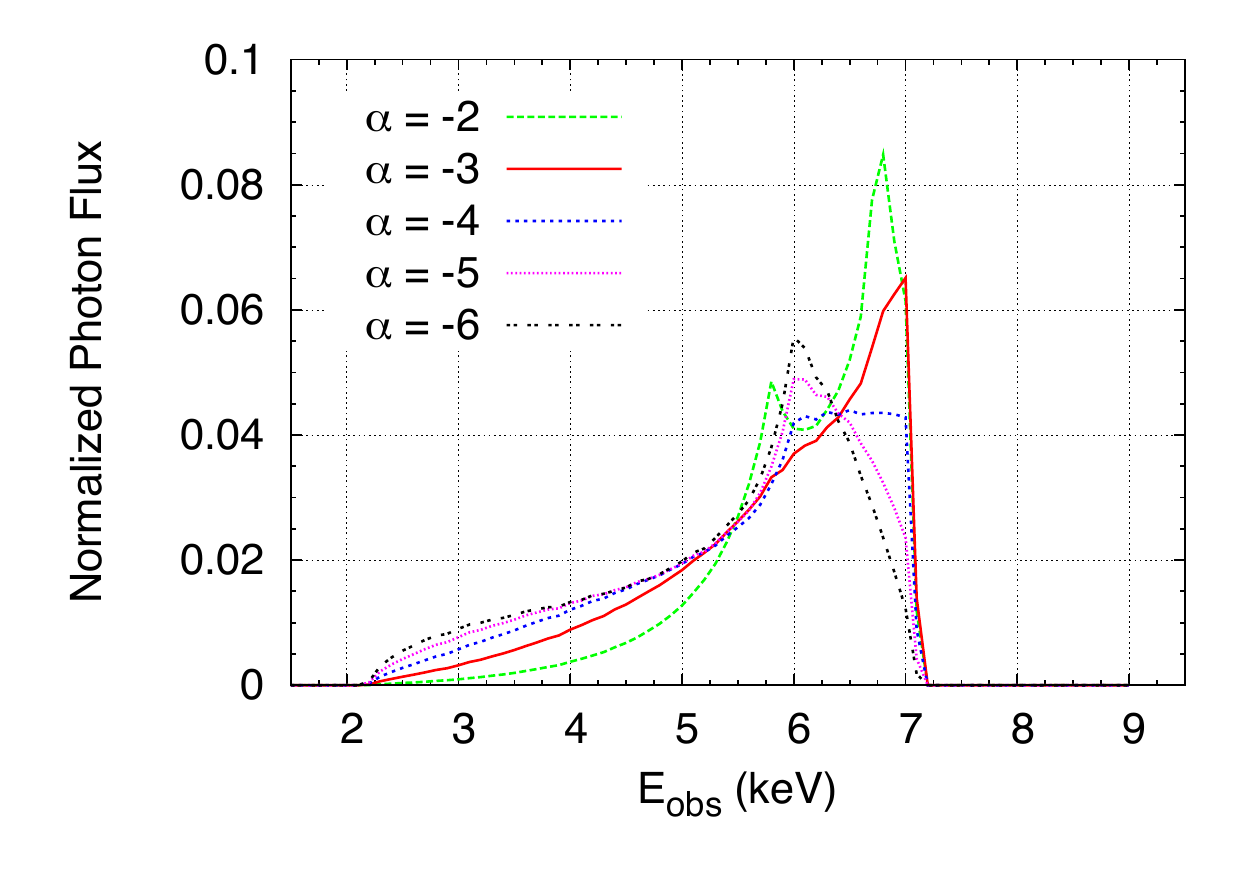} \\
\includegraphics[type=pdf,ext=.pdf,read=.pdf,width=8cm]{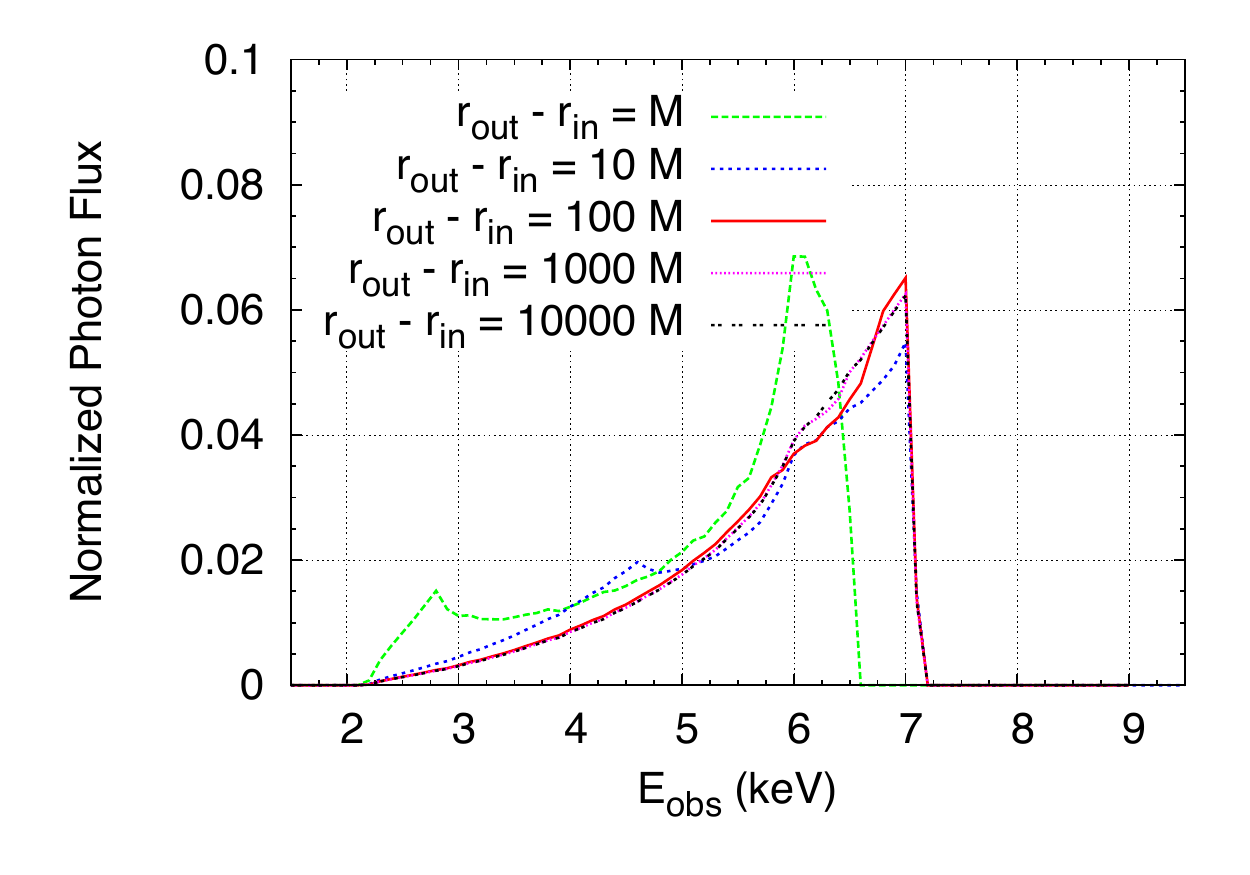} 
\includegraphics[type=pdf,ext=.pdf,read=.pdf,width=8cm]{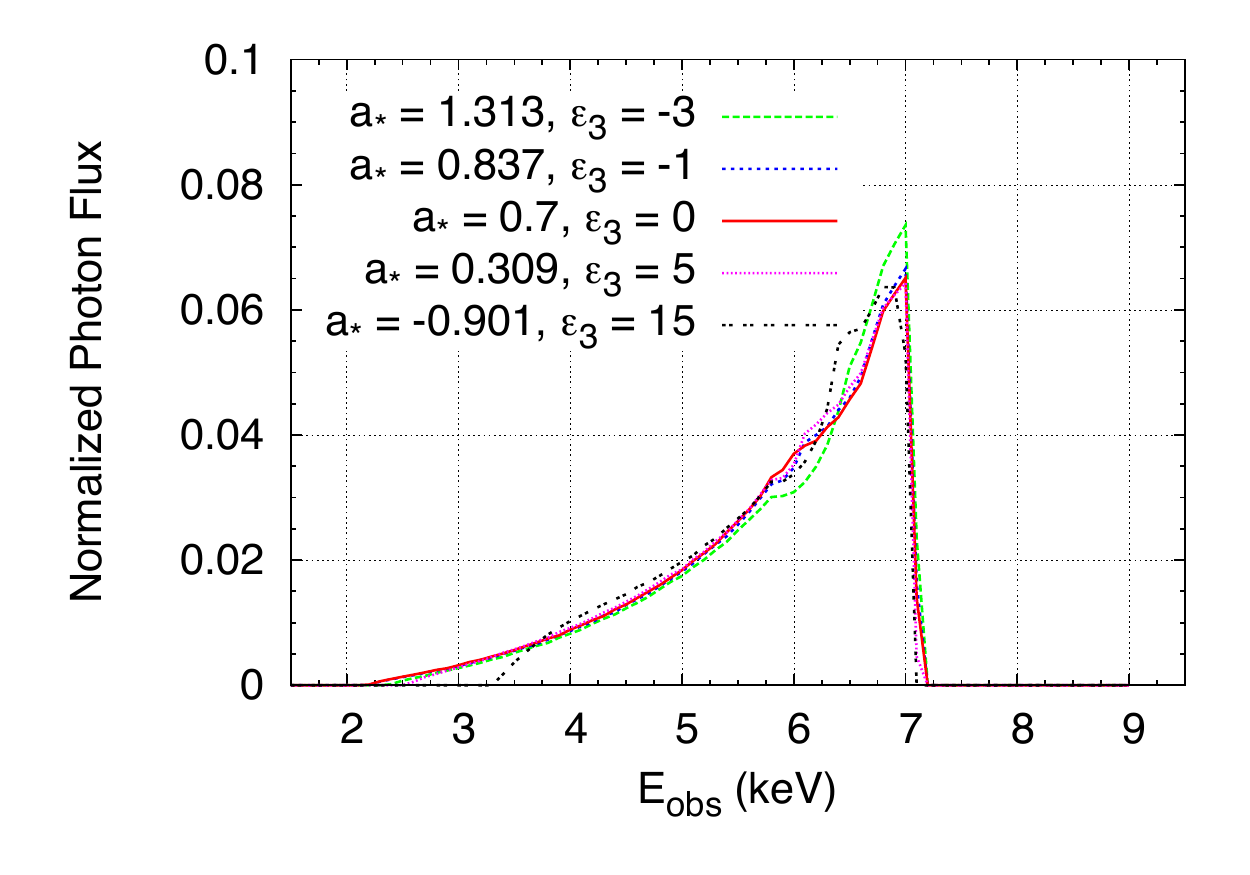}
\end{center}
\vspace{-0.6cm}
\caption{Broad K$\alpha$ iron line as a function of the parameters
of the model. Left top panel: spin parameter of the BH candidate.
Right top panel: deformation parameter $\epsilon_3$. Left central
panel: disk's inclination angle with respect to the line of sight of the 
distant observer. Right central panel: power-law index of the
intensity profile. Left bottom panel: size of the emission region. 
Right bottom panel: spin parameter and deformation parameter for
BH candidates with the same radiative efficiency $\eta = 0.1036$. 
When not specified, the value of the parameters is: 
$a_* = 0.7$, $\epsilon_3 = 0.0$,
$i = 45^\circ$, $\alpha = -3$, $r_{\rm out} - r_{\rm in} = 100$~$M$.}
\label{f-ka}
\end{figure}

\begin{figure}
\begin{center}  
\includegraphics[type=pdf,ext=.pdf,read=.pdf,width=8cm]{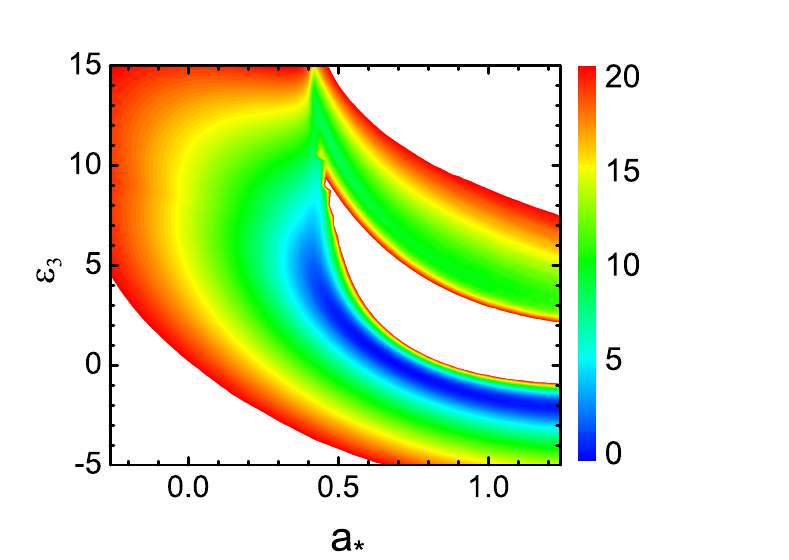}
\includegraphics[type=pdf,ext=.pdf,read=.pdf,width=8cm]{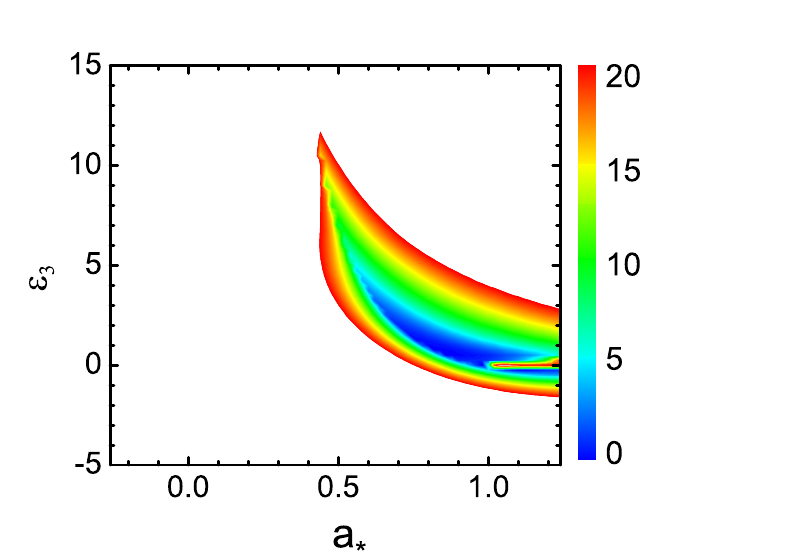}
\end{center}
\vspace{-0.6cm}
\caption{$\chi^2_{\rm red}$ from the comparison of the broad K$\alpha$ iron
line generated around a Kerr BH with spin parameter $\tilde{a}_*$ and a 
JP BH with spin parameter $a_*$ and deformation parameter $\epsilon_3$. 
Left panel: $\tilde{a}_* = 0.7$. Right panel: $\tilde{a}_* = 0.98$. See text for 
details.}
\label{f-ka_fit}
\vspace{1.0cm}
\begin{center}  
\includegraphics[type=pdf,ext=.pdf,read=.pdf,width=8cm]{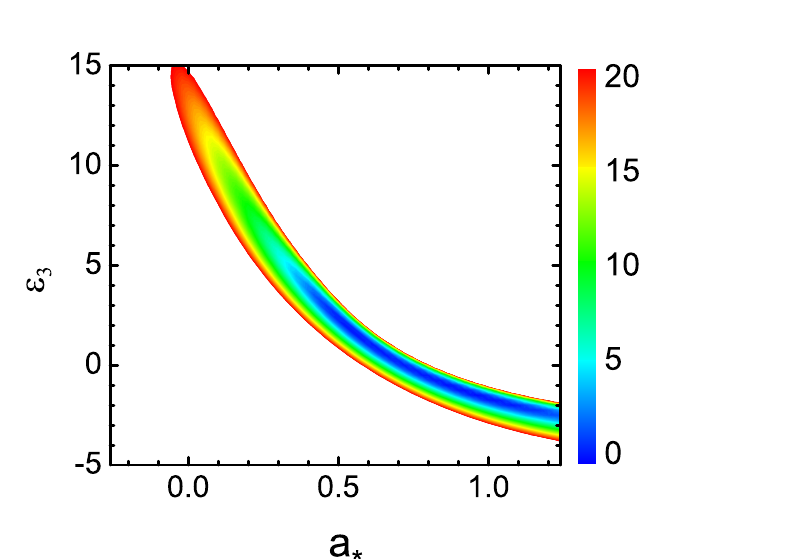}
\includegraphics[type=pdf,ext=.pdf,read=.pdf,width=8cm]{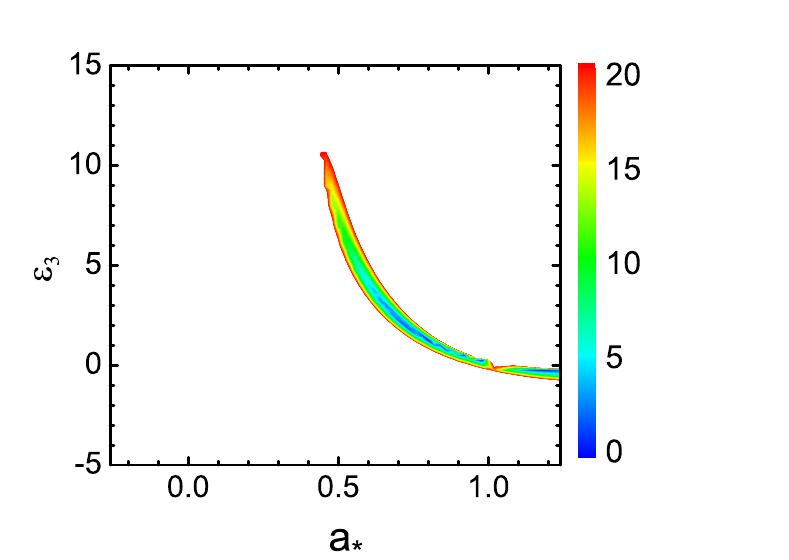}
\end{center}
\vspace{-0.6cm}
\caption{$\chi^2_{\rm red,tot}$ from the combination of the analysis of the 
broad K$\alpha$ iron line and the continuum-fitting method for the case
$\tilde{a}_* = 0.7$ (left panel) and $\tilde{a}_* = 0.98$ (right panel). See 
text for details.}
\label{f-kacfm}
\end{figure}

\subsection{Broad K$\alpha$ iron line}

The X-ray spectrum of both stellar-mass and super-massive BH candidates
is often characterized by the presence of a power-law component. This
feature is commonly interpreted as the inverse Compton scattering of 
thermal photons by electrons in a hot corona above the accretion disk.
The geometry of the corona is not known and several models have been
proposed. Such a ``primary component'' irradiates the accretion disk, 
producing a ``reflection component'' in the X-ray spectrum. The illumination of 
the cold disk by the primary component produces also spectral lines by 
fluorescence. The strongest line is the K$\alpha$ iron line at 6.4~keV. This 
line is intrinsically narrow in frequency, while the one observed appears 
broadened and skewed. The interpretation is that the line is strongly altered 
by special and general relativistic effects, which produce a characteristic 
profile, first predicted in \citet{fab89} and then observed for the first time 
in the ASCA data of the Seyfert~1 galaxy MCG-6-30-15 \citep{tan95}. In
MCG-6-30-15, this line is extraordinary stable, in spite of 
a substantial variability of the continuum, suggesting that the analysis of its 
shape can be used to probe the geometry of the space-time around the BH 
candidate. However, the relativistic origin of 
the observed broad K$\alpha$ iron lines is not universally accepted, and 
some authors have proposed different explanations \citep{turner,tita}.

Within the interpretation of a relativistically broadened K$\alpha$ iron line,
the shape of the line is primarily determined by the background metric, the 
geometry of the emitting region, the disk emissivity, and the disk's inclination 
angle with respect to the line of sight of the distant observer \citep{fab89,iron2,iron3}.
The effects of deviations from the Kerr metric on the shape of the K$\alpha$
iron line are discussed in \citet{jp-iron1}, \citet{jp-iron2}, and \citet{cb-iron}. 
In Kerr and non-Kerr backgrounds, $M$ sets the length of the system, so
everything scales as $M$ or as some power of $M$, without affecting the 
shape of the line. The relevant parameters of the background geometry
are thus the spin $a_*$ and possible non-vanishing deformation parameters.
In those sources for which there is indication that the K$\alpha$ iron line is 
mainly emitted close to the compact object, the emission region may be 
thought to range from the ISCO radius, $r_{\rm in} = r_{\rm ISCO}$, to some 
outer radius $r_{\rm out}$. However, even more complicated geometries have 
been proposed \citep{er}. In principle, the disk emissivity may be theoretically 
calculated. In practice, that is not feasible at present. The simplest choice is an 
intensity profile $I_{\rm e} \propto r^{\alpha}$, with $\alpha < 0$ a free parameter 
to be determined during the fitting procedure.

The line spectra can be expressed in terms of the photon flux number density as 
measured by the distant observer:
\be
N_{E_{\rm obs}} &=& \frac{1}{E_{\rm obs}} 
\int I_{\rm obs}(E_{\rm obs}) d \Omega_{\rm obs} =
\frac{1}{E_{\rm obs}} \int g^3 I_{\rm e}(E_{\rm e}) 
d \Omega_{\rm obs} \, .
\ee
Let us assume that the disk emission is 
monochromatic (the rest frame energy is $E_{\rm{K}\alpha} = 6.4$~keV) and 
isotropic with a power-law radial profile:
\be
I_{\rm e}(E_{\rm e}) \propto \delta (E_{\rm e} - E_{\rm{K}\alpha}) r^{\alpha} \, .
\ee
The resulting broad K$\alpha$ iron lines for different values of the 
model parameters (spin $a_*$, deformation $\epsilon_3$, inclination angle $i$, 
power-law index $\alpha$, and outer radius $r_{\rm out}$) are shown in 
Figure~\ref{f-ka}. The photon flux has been normalized so that
\be
\int N_{E_{\rm obs}} d E_{\rm obs} = {\rm constant} \, ,
\ee
as only the shape matters.

As we can see in Figure~\ref{f-ka}, the effect of variations of $a_*$ and of the 
deformation parameter $\epsilon_3$ is similar. For small deformations, backgrounds 
with the same efficiency in the NT model, $\eta_{\rm rad} = 1 - E_{\rm ISCO}$, 
have similar broad K$\alpha$ iron lines and it is difficult to
distinguish a Kerr BH from another compact object (see the right bottom panel).
However, unlike in the case of the thermal spectrum of a thin disk, for larger
deformations that is not true any more. The location of the
ISCO radius (set by $a_*$ and $\epsilon_3$ and determining $\eta_{\rm rad}$)
is encoded in the low-energy tail of the broad line. It consists of photons 
emitted from the receding part of the accretion disk close to the inner edge 
and thus strongly affected by both gravitational and Doppler redshift.
The main peak of the line is produced by the effect of Doppler 
blueshift, relatively far from the compact object, where the effect of gravitational
redshift is not so strong. For small values of the deformation
parameters, the Keplerian velocity profile is still similar to the one of a Kerr
BH and an object with similar radiative efficiency may look like a Kerr BH with
different spin parameter. For larger deviations from the Kerr background, the
role of the Doppler boosting (in both the main peak and the low-energy tail) 
becomes more and more important and breaks
the degeneracy between spin and deformation parameter. To be more
quantitative, we can proceed as in the previous section and compare
the line produced in the space-time around a Kerr BH and a deformed object.
The reduced $\chi^2$ is now:
\be\label{eq-chi2-ka}
\chi^2_{\rm red} (a_*, \epsilon_3, i, \alpha, r_{\rm out})
= \frac{\chi^2}{n} =
\frac{1}{n} \sum_{i = 1}^{n} \frac{\left[N_i^{\rm JP} 
(a_*, \epsilon_3, i, \alpha, r_{\rm out}) - N_i^{\rm Kerr}
(\tilde{a}_*, \tilde{i}, \tilde{\alpha}, \tilde{r}_{\rm out}) 
\right]^2}{\sigma^2_i} \, ,
\ee 
where the summation is performed over $n$ sampling 
energies $E_i$, $N_i^{\rm JP}$ and $N_i^{\rm Kerr}$ are the normalized photon 
fluxes in the energy bin $[E_i,E_i+\Delta E]$, respectively for the JP and Kerr metric,
and $\sigma_i$ is the uncertainty of the energy bin. Motivated by the fact that
variations of $i$, $\alpha$, and $r_{\rm out}$ produce different effects on the
shape of the line with respect to $a_*$ and $\epsilon_3$, for the sake of
simplicity we can fix $i = \tilde{i}$, $\alpha = \tilde{\alpha}$, and $r_{\rm out} = 
\tilde{r_{\rm out}}$ (actually, such an approximation may not be so good for the
inclination angle, see \citet{cb-iron}). Assuming a 10\% uncertainty in the photon count,
i.e. $\sigma_i = 0.1 \, N_i^{\rm Kerr}$, and data in the energy range $1 - 7.5$~keV
with an energy resolution $\Delta E=100$~eV, we get the plots in Figure~\ref{f-ka_fit}
for $\tilde{a} = 0.7$ (left panel) and 0.98 (right panel). Unlike for the analysis of 
the disk's spectrum, here we can exclude that a source with a K$\alpha$ iron line
that looks like the one generated in a Kerr space-time with spin parameter 
$\tilde{a} = 0.7$ comes instead from the region around a very deformed compact 
object.

It is also interesting to see what happens if we combine the analysis of the thermal 
spectrum of the disk with the one of the broad K$\alpha$ iron line. In general,
it is not common to have good measurements of both the disk's spectrum and the 
K$\alpha$ iron line for the same object, but it is possible and it may be more
likely with future experiments. If we define
\be
\chi^2_{\rm red,\, tot}  = 
\chi^2_{\rm red,\, K\alpha} + \chi^2_{\rm red,\, cfm} \, ,
\ee
where $\chi^2_{\rm red,\, K\alpha}$ and $\chi^2_{\rm red,\, cfm}$ are, respectively, 
the reduced $\chi^2$ defined in Eq.~(\ref{eq-chi2-ka}) and (\ref{eq-chi2}),
we find the two plots shown in Figure~\ref{f-kacfm}.

\begin{table}
\begin{center}
\begin{tabular}{c c c c c}
\hline \\
BH binary & $M/M_\odot$ & $\nu_{\rm U}$/Hz & $\nu_{\rm L}$/Hz & References \\ \\
\hline
GRO~J1655-40 & $6.30 \pm 0.27$ & $450 \pm 3$ & $300 \pm 5$ & \citet{strohmayer} \\
XTE~J1550-564 & $9.1 \pm 0.6$ & $276 \pm 3$ & $184 \pm 5$ & \citet{rem02} \\ 
GRS~1915+105 & $14.0 \pm 4.4$ & $168 \pm 3$ & $113 \pm 5$ & \citet{mcc} \\
\hline
\end{tabular}
\end{center}
\caption{Stellar-mass BH candidates in binary systems with a measurement of the 
mass and two observed high-frequency QPOs.}
\label{tab}
\end{table}

\subsection{Quasi Periodic Oscillations}

The X-ray power density spectra of some low-mass X-ray binaries show some 
peaks; that is, there are QPOs in the X-ray flux \citep{mcc}. These features can 
be observed in systems with both BH candidates and neutron stars and the 
frequencies of these oscillations are in the range 0.1~Hz--1~kHz. High-frequency 
QPOs in BH candidates ($\sim$40-450~Hz) are particularly interesting, as they 
depend only very weakly on the observed X-ray flux and for this reason it is 
thought they are determined by the metric of the space-time rather than by 
the properties of the accretion flow. If that is correct, they may be used to probe 
the geometry around stellar-mass BH candidates. For the time being, however, 
the exact physical mechanism responsible for the production of the high-frequency 
QPOs is not known and several different scenarios have been proposed, 
including hot-spot models~\citep{hsm1,hsm2}, diskoseismology 
models~\citep{dm1,dm2,dm3,dm4}, resonance models~\citep{rm1,rm2,ka05,rm3},
and p-mode oscillations of a small accretion torus~\citep{rezz1,rezz2,rezz3}. 
In these models, the frequencies of the QPOs are directly related to the three
characteristic orbital frequencies of a test-particle: the Keplerian frequency 
$\nu_{\rm K}$ (which is the inverse of the orbital period), the radial epicyclic 
frequency $\nu_r$ (the frequency of radial oscillations around the mean orbit), and 
the vertical epicyclic frequency $\nu_\theta$ (the frequency of vertical oscillations 
around the mean orbit). These three frequencies depend on the geometry of the 
space-time and on the radius of the orbit. While they are defined as the characteristic 
frequencies of the orbital motion for a free particle, there is a direct relation between 
these frequencies and the ones of the oscillation modes of the fluid accretion flow.

In the Kerr space-time, the radial epicyclic frequency $\nu_r$ reaches a maximum 
at some radius $r_{\rm max} > r_{\rm ISCO}$ and then vanishes at the ISCO, as
the orbit becomes marginally radially unstable. Keplerian and vertical epicyclic 
frequencies are instead defined up to the photon orbit. Circular orbits with radius
smaller than the one of the photon orbit do not exist. $\nu_\theta > \nu_r$, and for 
corotating orbits $\nu_{\rm K} \ge \nu_\theta$. If the space-time geometry around 
BH candidates is described by the Kerr metric, the measurement of the high-frequency 
QPOs may provide an estimate of the BH spin parameter, assuming the exact 
mechanism responsible for these oscillations is known. \citet{tkss} compare the 
available spin measurements from high-frequency QPOs and the continuum-fitting 
method and find that the two approach do not provide consistent results if we believe 
that all the observed high-frequency QPOs can be explained within a unique model.

The high-frequency QPO models developed for the Kerr metric can be
generalized to non-Kerr backgrounds \citep{jp-qpo,aliev,cb-qpo}. The possibility 
of the existence of vertically unstable orbits, absent in the Kerr space-time, 
allows for a larger number of possible combinations between different modes.
$\nu_\theta > \nu_r$ is not true any more. The conflict between 
measurements from high-frequency QPOs and the continuum-fitting method can 
be solved within a unique astrophysical model by admitting that the geometry 
around BH candidates deviates from the Kerr solution \citep{cb-qpo}.

In four stellar-mass BH candidates, we observe two high-frequency QPOs. It is 
remarkable that in all the four cases the ratio of the two frequencies is 3:2,
suggesting a strong correlation between them. Possible theoretical models
should thus be able to explain this feature. The resonance models
\citep{rm1,rm2,ka05,rm3} seem to be quite appealing from this point of view. 
For a free particle, the radial and the vertical modes are decoupled (see Eqs.~(\ref{eq-o1}) 
and (\ref{eq-o2}) in Appendix). However, it is natural to expect that in a 
more realistic description there are non-linear effects coupling the two epicyclic 
modes. In this case, the equations can be written as 
\be\label{eq-o3}
\frac{d^2 \delta_r}{dt^2} + \Omega_r^2 \delta_r &=& 
\Omega_r^2 F_r \left( \delta_r, \delta_\theta, \frac{d \delta_r}{dt}, 
\frac{d \delta_\theta}{dt}\right) \, , \\
\frac{d^2 \delta_\theta}{dt^2} + \Omega_\theta^2 \delta_\theta &=& 
\Omega_\theta^2 F_\theta \left( \delta_r, \delta_\theta, \frac{d \delta_r}{dt}, 
\frac{d \delta_\theta}{dt}\right) \, ,
\ee
where $F_r$ and $F_\theta$ are some functions that depend on the specific 
properties of the accretion flow. If we knew the details of the physical
mechanisms of the accretion process, we could write the explicit form of these
two functions and solve the system. Unfortunately, that is not the case.
However, regardless of the specific resonance model and the microphysics
responsible for it, the resonance paradigm requires that the upper and the 
lower frequencies have the form
\be
\nu_{\rm U} &=& m_1 \nu_r + m_2 \nu_\theta \, , \\
\nu_{\rm L} &=& n_1 \nu_r + n_2 \nu_\theta \, . 
\ee
where $m_1$, $m_2$, $n_1$, and $n_2$ are integer (and likely as small as 
possible) numbers. Scenarios with a coupling between $\nu_r$ and 
$\nu_\theta$ are surely theoretically more motivated, but it is not possible to
exclude {\it a priori} that $\nu_{\rm U}$ and $\nu_{\rm L}$ actually arise from
the combination of $\nu_{\rm K}$ with one of the other two frequencies and
have the form like $\nu_{\rm U} = m_1 \nu_{\rm K} + m_2 \nu_r$ and $\nu_{\rm L} 
= n_1 \nu_{\rm K} + n_2 \nu_r$. Assuming the JP background with deformation 
parameter $\epsilon_3$, we can compare the theoretical predictions with the observed
$\nu_{\rm U}$ and $\nu_{\rm L}$ for any BH candidate and define an allowed 
region on the plane spin parameter-deformation parameter for any specific set
of $\{ m_1, m_2, n_1, n_2 \}$.

While systematic effects and/or wrong models are surely the most likely 
possibility to explain the disagreement between the spin measurements
obtained by the continuum-fitting method and the high-frequency QPOs,
\citet{cb-qpo} explores the possibility of explaining all the observations with
a unique astrophysical model in a non-Kerr background.
It turns out that the resonance $\nu_\theta : \nu_r = 3:1$ can do the job
if the JP deformation parameter $\epsilon_3$ is in the range 5 to 15.
In this scenario, GRO~J1655-40 and XTE~J1550-564 would be ``slow-rotating'' 
compact objects and their ISCO would be marginally radially stable.
GRS~1915+105 would instead be a ``fast-rotating'' object with $a_* \approx 0.5$
and its ISCO would be marginally vertically stable.

\begin{figure}
\begin{center}  
\includegraphics[type=pdf,ext=.pdf,read=.pdf,width=8cm]{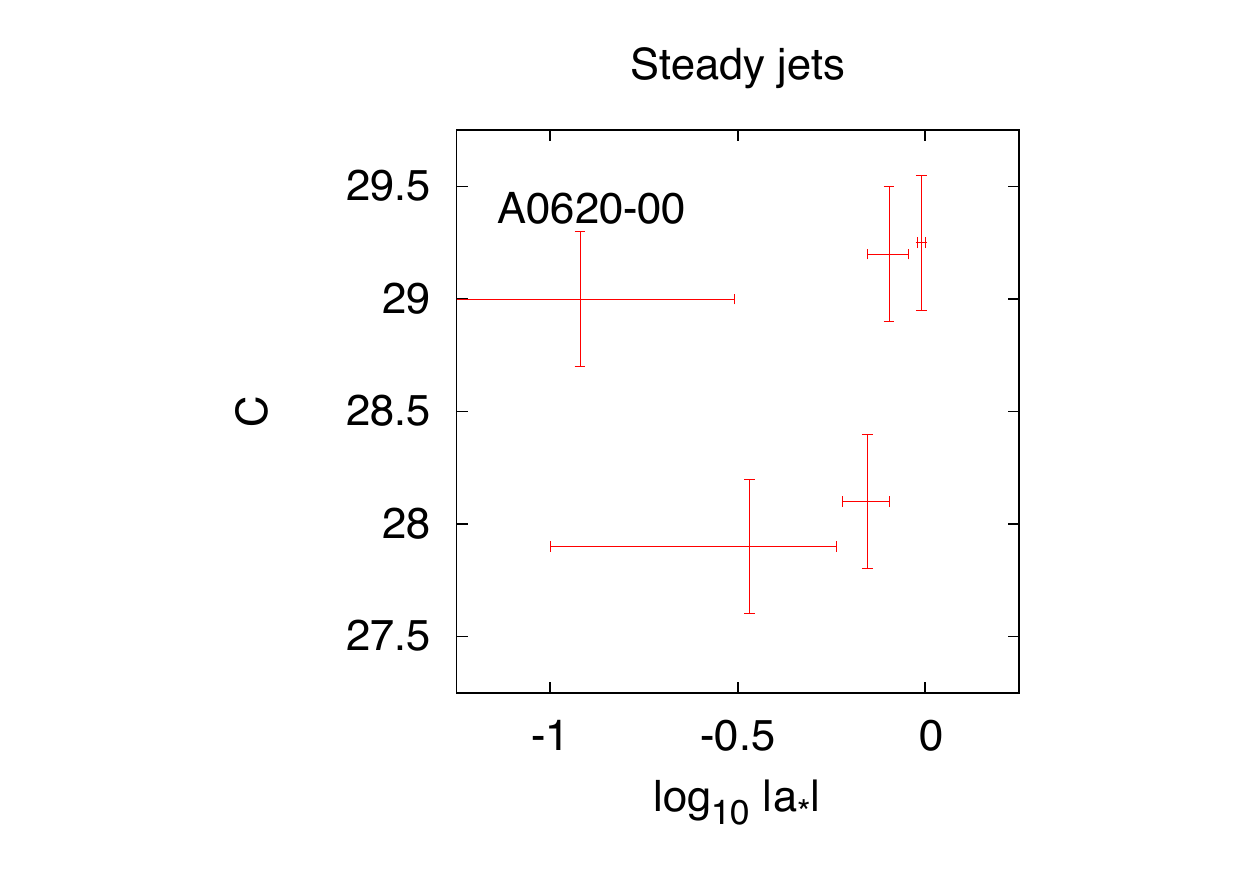}
\includegraphics[type=pdf,ext=.pdf,read=.pdf,width=8cm]{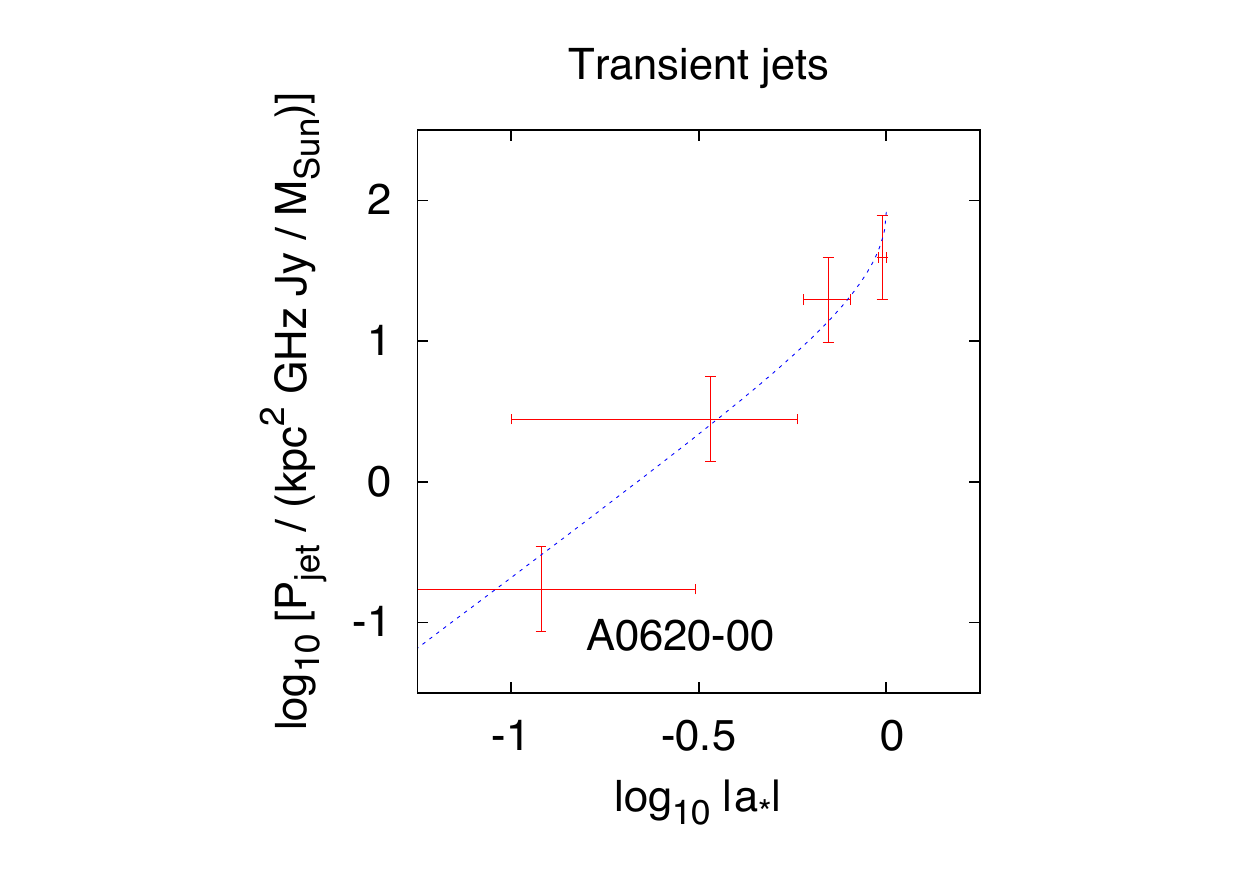}
\end{center}
\vspace{-0.6cm}
\caption{Left panel: absence of evidence for a correlation between the
jet power and the BH spin for steady jets \citep{fgr}. Right panel: evidence 
for a correlation between the jet power and the BH spin for transient jets
\citep{nm}. See text for details.}
\label{f-jet-k}
\vspace{1.0cm}
\begin{center}  
\includegraphics[type=pdf,ext=.pdf,read=.pdf,width=8cm]{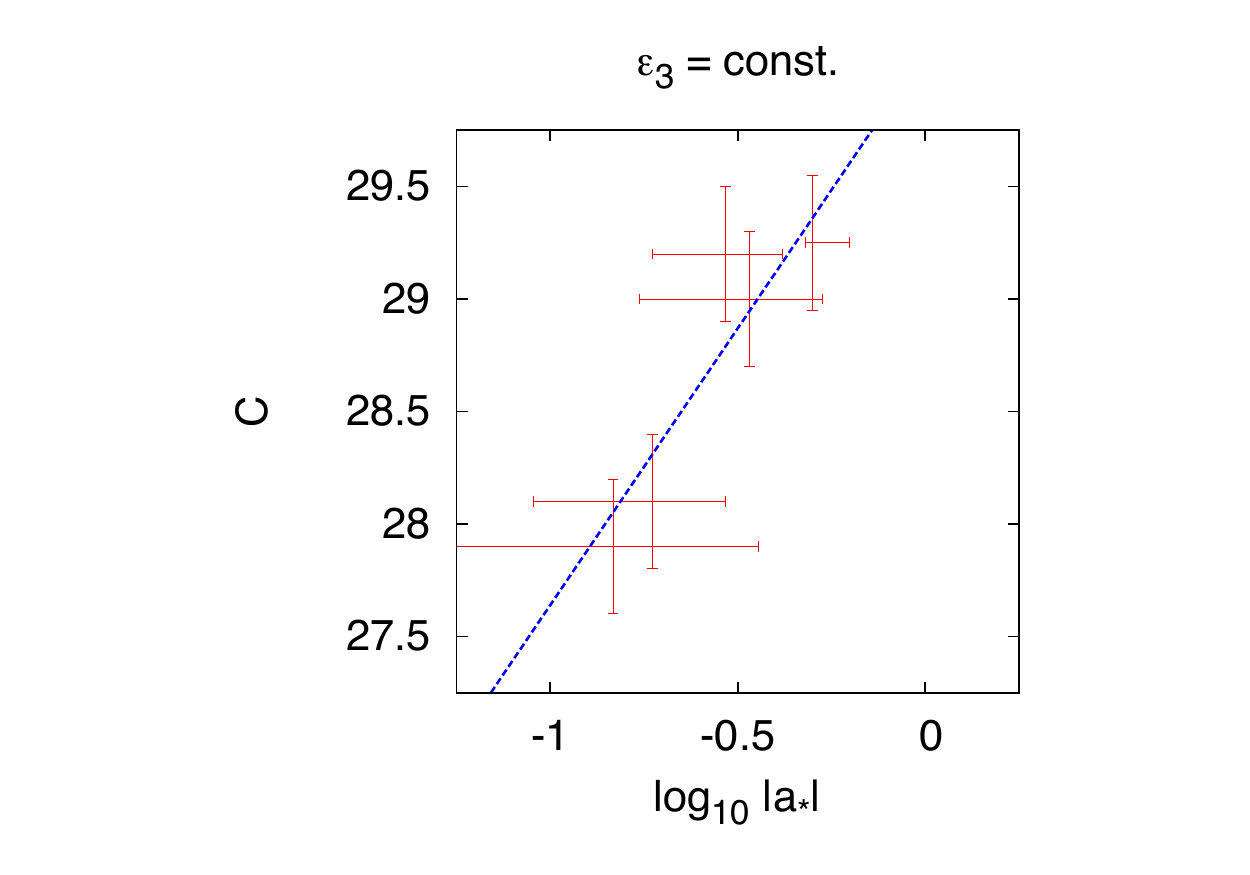}
\includegraphics[type=pdf,ext=.pdf,read=.pdf,width=8cm]{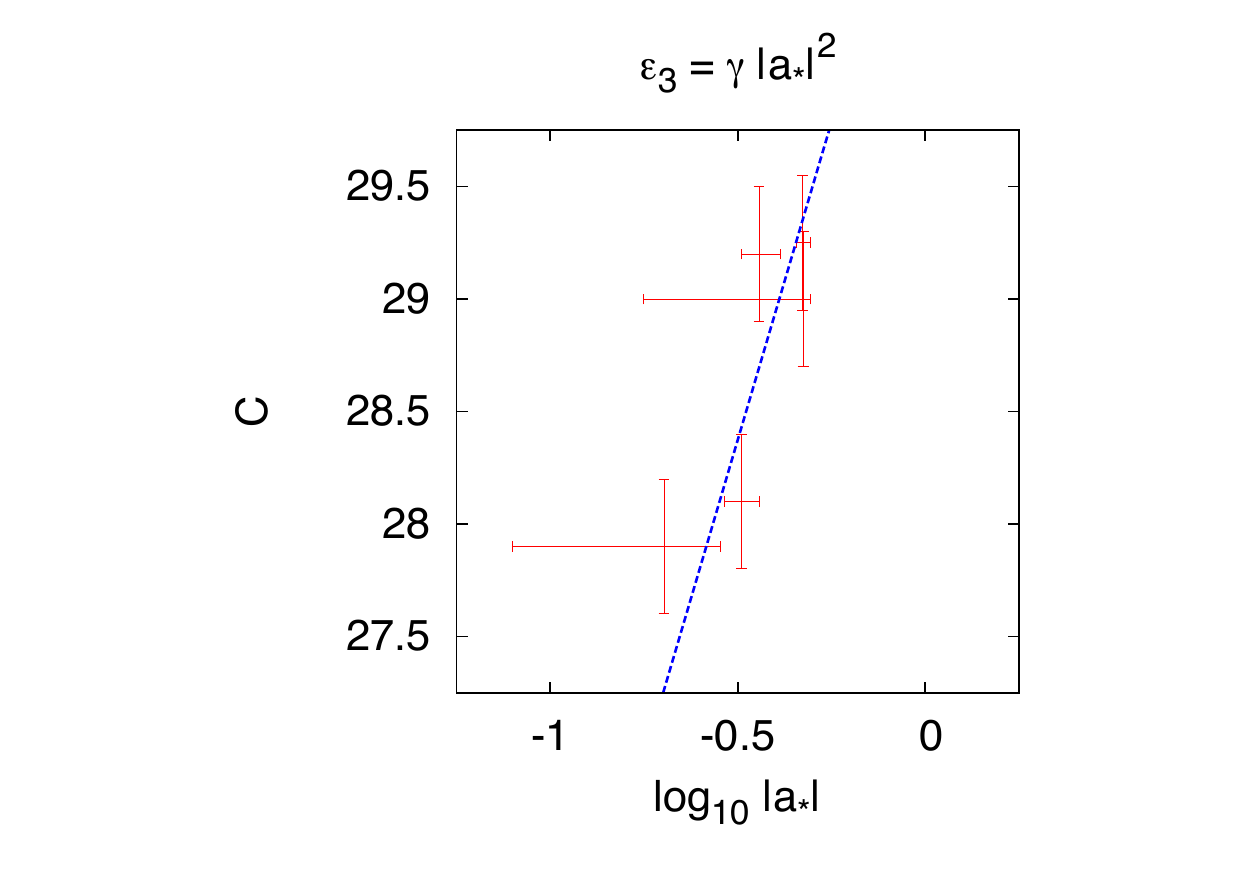}
\end{center}
\vspace{-0.6cm}
\caption{Best fit for a possible correlation between the jet power and the
spin parameter of BH candidates assuming a non-vanishing deformation 
parameter $\epsilon_3$. Left panel: $\epsilon_3$ constant for all the 
objects; the best fit is for $\epsilon_3 = 7.5$. Right panel: $\epsilon_3 = 
\gamma |a_*|^2$, with $\gamma$ constant; the best fit is for $\gamma = 45$.
See \citet{cb-jet2} for more details.}
\label{f-jet-jp}
\end{figure}

\subsection{Jet power}

Jets and outflows are a quite common feature of accreting compact objects.
In the case of stellar-mass BH candidates in X-ray binary systems, we 
observe two kinds of jets~\citep{fbg}. {\it Steady jets} occur in the hard spectral 
state. {\it Transient jets} appear most significantly when the source switches 
from the hard to the soft state. Most efforts so far concentrate on the formation 
mechanism of steady jets. One of the most appealing scenarios to explain 
the formation of steady jets is the Blandford-Znajek mechanism~\citep{bz}, 
in which magnetic fields threading the BH event horizon are twisted and 
can extract the rotational energy of the spinning BH, producing an electromagnetic 
jet. Numerical simulations show that the mechanism can be very efficient 
and depends strongly on the BH spin~\citep{mck,tch1,tch2}. At the moment it 
is not clear if the Blandford-Znajek mechanism can be responsible for the 
production of steady jets, and in the literature there are some controversial 
results. The spin scenario is surely attractive, but still unproved.

Recently, there have been some studies investigating if there is observational 
evidence for a correlation between BH spin and jet power in current data.
\citet{fgr} consider all the spin measurements of BH binaries reported in the 
literature from the continuum-fitting method and the K$\alpha$ 
iron line analysis. For steady jets in the hard spectral state, they estimate the 
jet power via a normalization $C$, defined by
\be
\log_{10} L_{\rm radio} = C + 0.6 \left( \log_{10} L_{\rm X} - 34 \right) \, ,
\ee
where $L_{\rm radio}$ and $L_{\rm X}$ are, respectively, the radio and X-ray
luminosity of the object. Their plots show no evidence for a correlation 
between BH spin and jet power. The left panel of Figure~\ref{f-jet-k} shows the
case in which the spin parameter is measured with the continuum-fitting method. 
If the measurements of the spins and of the jet powers are correct, it seems
that steady jets are not powered by the BH spin.

\citet{nm} propose that the Blandford-Znajek mechanism is responsible for the 
formation of transient jets. They consider the most recent spin measurements of the 
continuum-fitting method from the Harvard-Smithsonian CfA group, which are supposed 
to be more reliable, and use a different proxy for the jet power, the peak radio 
flux normalized at 5~GHz, which they claim to be model independent. These 
authors find a correlation between BH spin $a_*$ and jet power $P_{\rm jet}$ 
(see the right panel in Figure~\ref{f-jet-k}), which is consistent with both the 
theoretical prediction
\be\label{eq-bz-p}
P_{\rm jet} \propto a_*^2
\ee
obtained in \citet{bz} for $a_*^2 \ll 1$ and the more accurate one 
\be
P_{\rm jet} \propto \Omega_{\rm H}^2 \propto a_*^2/r_{\rm H}^2 \, ,
\ee 
where $\Omega_{\rm H}$ and $r_{\rm H}$ are, respectively, the angular frequency 
and the radius of the event horizon, found in \citet{tch1} and valid even when 
$a_*$ is quite close to 1. However, numerical studies of the Blandford-Znajek 
mechanism show the production of steady jets, while the origin of transient jets 
remains unclear and other scenarios may look more promising, such as episodic 
ejections of plasma blobs~\citep{fy}.

In \citet{fgr} and \citet{nm}, the authors rely on spin measurements obtained 
under the assumption that BH candidates are Kerr BHs. If this hypothesis were
wrong, all the spin measurements would be wrong too. One the contrary, a jet powered 
by the Blandford-Znajek mechanism should still be correlated with the spin of 
the compact object. The existence of the event horizon is not strictly necessary
for the production of jets powered by the spin: for example, even neutron stars 
may have spin powered jets if there are magnetic fields anchored on the neutron star. 
On very general arguments, if the jet is powered by the rotational energy of
the compact object, the power can be written as
\be\label{eq-th-p}
P_{\rm jet} = A_0 + 
\sum_{n = 1}^{+\infty} A_n |a_*|^{2n} \, ,
\ee
where $A_0$ takes into account a possible non-spin contribution ($ A_0 \ge 0$, 
as $P_{\rm jet}$ cannot become negative for $a_* = 0$) and the other terms are 
due to (some version of) the Blandford-Znajek mechanism. The latter must 
depend only on even powers of $a_*$, because the direction of the spin should 
not be important (at least at first approximation).

Let us now examine the spin measurements obtained with the continuum-fitting
method, as they can be easily translated into efficiency measurements in the
NT model, and then into allowed regions on the spin parameter-deformation 
parameter plane. From Section~\ref{s-cfm}, one can see that an object more
prolate than a Kerr BH ($\epsilon_3 > 0$) may be interpreted as a Kerr BH with 
a higher value of the spin parameter. An object more oblate than a Kerr BH 
($\epsilon_3 < 0$) has instead a spectrum similar to a Kerr BH with a lower value of 
$a_*$. While the jet power is independent of the spin sign, the continuum-fitting 
method is not, and the measurement of a very low radiative efficiency 
can be attributed to an object with a retrograde accretion disk.

If we believe that steady jets are powered by the spin of the compact object,
we can see if a correlation between jet power and spin parameter is possible
for a non-vanishing deformation parameter \citep{cb-jet2}. In the left panel of 
Figure~\ref{f-jet-k}, the absence of a correlation between $a_*$ and $P_{\rm jet}$ is 
essentially due to the powerful steady jet of A0620-00, while the continuum-fitting method
would suggest (for a Kerr metric) that this is a slow-rotating object. If astrophysical 
BH candidates were objects more prolate than Kerr BHs, their spin parameter
would be lower than the one reported in Table~\ref{tab1}. In particular, for a sufficiently 
large deformation parameter, A0620-00 can be consistent with a fast-rotating
object with a retrograde accretion disk. In this case, $P_{\rm jet}$ could be high, 
and a correlation between power of steady jets and spin parameter is possible.
If we neglect a possible non-spin contribution to the power of the jet, a simple 
form for $P_{\rm jet}$ is
\be\label{eq-power}
P_{\rm jet} = \alpha |a_*|^\beta \, ,
\ee
and therefore
\be
C = \beta \log_{10}|a_*| + \alpha'
\ee
where $\alpha' = \log_{10}\alpha$ and $\beta$ are the two parameters of the 
jet model. If we consider the JP metric with $\epsilon_3$ a universal constant 
common to all the objects, we find the best fit for $\epsilon_3 = 7.5$
$\alpha' = 30.1$, and $\beta = 2.46$ (left panel in 
Figure~\ref{f-jet-jp}). If we think it is physically more plausible a deformation 
parameter that depends on the features of the compact object, the simplest 
guess is likely
\be\label{eq-gamma}
\epsilon_3 = \gamma |a_*|^2 \, ,
\ee
with $\gamma$ constant. In this case, the best fit is for $\gamma = 45$, 
$\alpha' = 31.2$, and $\beta = 5.65$ (right panel in Figure~\ref{f-jet-jp}). 
The puzzling result in \citet{fgr} may thus be interpreted as an indication
that BH candidates are not the Kerr BHs predicted by General Relativity \citep{cb-jet2}.

If we instead believe that the Blandford-Znajek mechanism is responsible 
for the production of transient jets, as suggested in \citet{nm}, we can constrain 
possible deviations from the Kerr geometry \citep{cb-jet1}. Now A0620-00 shows a quite 
weak transient jet, which is indeed what one should expect for a slow-rotating 
object. If we consider objects more and more prolate than Kerr BHs, eventually,
on the base of the analysis of the thermal spectrum of its disk, A0620-00 should 
be interpreted as a fast-rotating object with a retrograde accretion disk. This 
would spoil the correlation between $a_*$ and $P_{\rm jet}$, as the Blandford-Znajek 
mechanism would predict a powerful jet for a fast-rotating object.

For the time being we do not know if steady or transient jets
are powered by the spin of the compact object and therefore we cannot really
use the estimate of the jet power to test the nature of BH candidates. If we believe 
that the Blandford-Znajek mechanism is responsible for the production of 
steady jets, as it is often thought, current observations would suggest that 
astrophysical BH candidates are not Kerr BHs. If we believe in the proposal of 
\citet{nm}, in which the Blandford-Znajek mechanism is responsible for the 
formation of transient jets, despite this is not the usual scenario, we can put a 
constraint on possible deviations from the Kerr metric. The Blandford-Znajek 
mechanism can unlikely be responsible for the creation of both steady and transient 
jets, even assuming an {\it ad hoc} deformation parameter for each object, as 
A0620-00 shows a quite powerful steady jet and a weak transient jet, which 
would require, respectively, either a high or a low value of the spin parameter.

\section{Conclusions}

Today we think the final product of the gravitational collapse is a Kerr BH
and we have identified several astrophysical candidates. The latter are
dark objects either in X-ray binary systems or in galactic nuclei, which are 
too heavy and compact to be, respectively, neutron stars or clusters of 
neutron stars. These two classes of objects are thought to be the Kerr BHs
predicted by General Relativity because there is no alternative explanation
in the framework of conventional physics. However, we have only robust 
measurements of the masses of these objects, while there is no clear
indication that the geometry of the space-time around them is really described 
by the Kerr metric.

The Kerr-nature of astrophysical BH candidates may be tested with future
space-based gravitational wave facilities \citep{ryan,gb06,bc07}, with the
possible discovery of a BH binary with a pulsar companion \citep{wex},
or by studying the orbital motion of new stars at milliparsec distance from the
super-massive BH candidate at the center of the Galaxy \citep{will08}.
However, all these data are not available now and it is definitively not
clear and easy to predict when they will be available in the future.
Recently, there have been an increasing interest in the possibility of
testing the geometry of the space-time around astrophysical BH candidates
by studying the properties of the electromagnetic radiation emitted by 
the gas in the accretion disk. These data are in most cases already available 
and, thanks to the significant progresses in the last $\sim 5$ years, people
have started using these data to extract information on the space-time
geometry around BH candidates.

In this paper, I review the motivations, a possible approach, and some physical
phenomena to probe the space-time geometry around BH candidates and
check if they are the Kerr BHs predicted by General Relativity. The continuum-fitting 
method and the analysis of the K$\alpha$ iron line are relatively mature 
techniques. They are now commonly used by astronomers to estimate the
spin parameter of these objects under the assumption of the Kerr background
and they can be naturally generalized to measure possible deviations from
the Kerr solution. In general, there is a strong degeneracy between the spin
and the deformation parameter; that is, it is not possible to measure the spin
and the deformation parameter at the same time, but only some combination
of them. That is particularly true in the case of the continuum-fitting method.
This degeneracy can be broken (or partially solved) if we can constrain
the space-time geometry around a specific object with two or more techniques.
The analysis of high frequency QPOs and of the jet power are other two 
interesting approaches, but the associated physical phenomena are not yet well 
understood and therefore, at present, they cannot really be used to test the Kerr-nature 
of BH candidates. In the case of the high frequency QPOs, in a Kerr background
there is no way to explain all the observations with a unique scenario and find
results consistent with the continuum-fitting method.
While this fact might be interpreted as a possible indication of new physics,
it is definitively possible that we do not have yet the correct model to
describe these phenomena or that there are systematics effects not properly
taken into account. In the case of the analysis of the jet power, we do not
know if steady or transient jets are powered by the spin of BH candidates.
The Blandford-Znajek mechanism can explain steady jets, but a correlation
between spin measurement and jet power in current data would require
that the space-time around BH candidates has deviation from the Kerr
solution. If we believe that the Blandford-Znajek mechanism is responsible
for the production of transient jets, despite this is not the usual scenario,
observations would be consistent with the interpretation that BH candidates
are Kerr BHs and we can put a bound on possible deviations from the Kerr 
geometry.

In the near future ($\lesssim 5$~years), it is also expected that we will be
able to observe the direct image of the super-massive BH candidate at the
center of the Galaxy with a resolution comparable to its gravitational radius
\citep{shadow}. In the case of a geometrically thick and optically thin accretion 
disk, we will observe the BH shadow, a dark area over a bright
background, whose shape depends only on the geometry of the space-time
around the compact object and it can thus be used to test the Kerr BH hypothesis 
\citep{cb-k,cb-n,jp-rt,cb-fl,cn2}.


\begin{acknowledgments}
I would like to thank Zilong Li for help in the preparation of the manuscript.
This work was supported by the Thousand Young Talents Program
and Fudan University.
\end{acknowledgments}


\appendix

\section{Circular orbits on the equatorial plane and ISCO}

The line element of a generic stationary and axisymmetric
space-time can be written as
\be
ds^2 = g_{tt} dt^2 + 2g_{t\phi}dt d\phi + g_{rr}dr^2 
+ g_{zz} dz^2 + g_{\phi\phi}d\phi^2 \, .
\ee
Since the metric is independent of the $t$ and $\phi$
coordinates, we have the conserved specific energy at
infinity, $E$, and the conserved $z$-component of the
specific angular momentum at infinity, $L_z$. This fact
allows to write the $t$- and $\phi$-component of the 
4-velocity of a test-particle as 
\be
\dot{t} = \frac{E g_{\phi\phi} + L_z g_{t\phi}}{
g_{t\phi}^2 - g_{tt} g_{\phi\phi}} \, , \qquad 
\dot{\phi} = - \frac{E g_{t\phi} + L_z g_{tt}}{
g_{t\phi}^2 - g_{tt} g_{\phi\phi}} \, .
\ee
From the conservation of the rest-mass, $g_{\mu\nu}\dot{x}^\mu \dot{x}^\nu = -1$,
we can write
\be
g_{rr}\dot{r}^2 + g_{zz}\dot{z}^2 = V_{\rm eff}(r,z) \, ,
\ee
where the effective potential $V_{\rm eff}$ is given by
\be
V_{\rm eff} = \frac{E^2 g_{\phi\phi} 
+ 2 E L_z g_{t\phi} + L^2_z g_{tt}}{
g_{t\phi}^2 - g_{tt} g_{\phi\phi}} - 1 \, .
\ee
Circular orbits in the equatorial plane are located at the zeros
and the turning points of the effective potential: 
$\dot{r} = \dot{z} = 0$, which implies $V_{\rm eff} = 0$,
and $\ddot{r} = \ddot{z} = 0$, requiring respectively 
$\partial_r V_{\rm eff} = 0$ and $\partial_z V_{\rm eff} = 0$.
From these conditions, one can obtain the angular velocity,
$E$, and $L_z$ of the test-particle:
\be
\Omega_\pm &=& \frac{d\phi}{dt} = 
\frac{- \partial_r g_{t\phi} 
\pm \sqrt{\left(\partial_r g_{t\phi}\right)^2 
- \left(\partial_r g_{tt}\right) \left(\partial_r 
g_{\phi\phi}\right)}}{\partial_r g_{\phi\phi}} \, , \\
E &=& - \frac{g_{tt} + g_{t\phi}\Omega}{
\sqrt{-g_{tt} - 2g_{t\phi}\Omega - g_{\phi\phi}\Omega^2}} \, , \\
L_z &=& \frac{g_{t\phi} + g_{\phi\phi}\Omega}{
\sqrt{-g_{tt} - 2g_{t\phi}\Omega - g_{\phi\phi}\Omega^2}} \, ,
\ee
where the sign $+$ is for corotating orbits and the sign $-$
for counterrotating ones.
The orbits are stable under small perturbations if 
$\partial_r^2 V_{\rm eff} \le 0$ and $\partial_z^2 V_{\rm eff} \le 0$.
In Kerr spacetime, the second condition is always satisfied,
so one can deduce the radius of the ISCO from $\partial_r^2 V_{\rm eff} = 0$.
In general, however, one has to check also the stability along the
vertical direction, as well as that the plunging region connects the
ISCO to the surface/horizon of the compact object \citep{cb-e}.

The radial and vertical epicyclic frequencies can be quickly computed 
by considering small perturbations around circular equatorial orbits, 
respectively along the radial and vertical direction. If $\delta_r$ and 
$\delta_\theta$ are the small displacements around the mean orbit 
(i.e. $r = r_0 + \delta_r$ and $\theta = \pi/2 + \delta_\theta$), we find they 
are governed by the following differential equations
\be\label{eq-o1}
\frac{d^2 \delta_r}{dt^2} + \Omega_r^2 \delta_r &=& 0 \, , \\
\frac{d^2 \delta_\theta}{dt^2} + \Omega_\theta^2 \delta_\theta &=& 0 \, ,
\label{eq-o2}
\ee
where
\be\label{eq-or}
\Omega^2_r &=& - \frac{1}{2 g_{rr} \dot{t}^2} 
\frac{\partial^2 V_{\rm eff}}{\partial r^2} \, , \\
\Omega^2_\theta &=& - \frac{1}{2 g_{\theta\theta} \dot{t}^2} 
\frac{\partial^2 V_{\rm eff}}{\partial \theta^2} \, .
\label{eq-ot}
\ee
The radial epicyclic frequency is $\nu_r = \Omega_r/2\pi$ and 
the vertical epicyclic frequency is $\nu_\theta = \Omega_\theta/2\pi$.


\end{document}